\documentclass[10pt,twoside,english,preprint,tightenlines]{revtex4-1}
\usepackage{lmodern}

\usepackage[T1]{fontenc}
\usepackage[latin9]{inputenc}
\usepackage{graphicx}
\usepackage{esint}
\usepackage{fullpage}

\makeatletter

\@ifundefined{textcolor}{}
{%
 \definecolor{BLACK}{gray}{0}
 \definecolor{WHITE}{gray}{1}
 \definecolor{RED}{rgb}{1,0,0}
 \definecolor{GREEN}{rgb}{0,1,0}
 \definecolor{BLUE}{rgb}{0,0,1}
 \definecolor{CYAN}{cmyk}{1,0,0,0}
 \definecolor{MAGENTA}{cmyk}{0,1,0,0}
 \definecolor{YELLOW}{cmyk}{0,0,1,0}
 }

\@ifundefined{definecolor}{\usepackage{color}}{}
\usepackage{ifpdf}
\ifpdf
\IfFileExists{lmodern.sty}{\usepackage{lmodern}}{}

\fi
\let\myTOC\tableofcontents\renewcommand{\tableofcontents}{%
  \frontmatter
  \pdfbookmark[1]{\contentsname}{}
  \myTOC
  \mainmatter }

\def\LyX{\texorpdfstring{%
  L\kern-.1667em\lower.25em\hbox{Y}\kern-.125emX\@}
  {LyX}}

\usepackage{babel}

\makeatother

\usepackage{babel}
\begin{document}

\title{The influence of water interfacial potentials on ion hydration free energies and density profiles near the surface}

\author{Thomas L. Beck}
\address{Departments of Chemistry and Physics  
University of
Cincinnati, Cincinnati, OH 45221-0172}

\begin{abstract}
The surface or contact potential at the water liquid-vapor interface is discussed in relation to determinations of absolute ion hydration free energies and distributions of ions near the interface.   It is shown that, rather than the surface potential itself, the net electrostatic potential at the center of an uncharged solute can aid both in relating differences between tabulations of hydration free energies and in explaining differing classical and quantum surface potential estimates.  Quantum mechanical results are consistent with conclusions from classical simulations that there is a net driving force that enhances anion density at the surface relative to cations.
\end{abstract}

\maketitle

\section{Introduction}

Both the sign and the magnitude of the surface potential $\phi_{sp}$ across
the water liquid/vapor interface have eluded experimental
and theoretical consensus, although recent progress has been made. 
Electron holography experiments suggest a surface potential of $+3.5$ V for vitrified ice \cite{holography}. Two density functional theory (DFT)
studies have obtained values of $+3.1$ V \cite{kathmann_understanding_2011} and $+3.6$ V \cite{leung_surface_2010} for the water liquid-vapor interface, consistent with the experimental result.  A related DFT study revealed a shift in ion HOMO states in clusters compared with bulk periodic boundary calculations, suggesting a surface potential value of $+4$ V \cite{sprik-sp-05}. 
Such a large potential change across a relatively narrow interface would naively suggest enormous
electric fields that would have substantial chemical effects.  Surface potentials computed with classical point-charge or polarizable models, on the other hand, are typically in the range $-0.6$ to $-0.5$ V \cite{tildesley_wsp,warren_hydration_2007,wick_effect_2007,harder_origin_2008,tlbsurf}. (The quoted values are for the SPC/E, TIP3P, TIP4P, Dang-Chang (polarizable) and TIP4P-FQ (polarizable) water models; recent work has shown that the surface potential of the TIP5P water model is closer to $-0.1$ V \cite{mundy-foot1}.)


It is generally accepted that the surface potential is not directly accessible to thermodynamic measurement \cite{guggenheim}. 
Nonetheless, electrochemical measurements have been used to 
indirectly infer surface potential values \cite{frumkin,farrell,trasatti,rusanov,jarvis,fawcett08} with magnitudes much smaller than 3.5 V,
ranging from $-0.4$ V to $+0.3$ V, as summarized by Randles \cite{randles77}.  
The experiments have employed a variety of approaches, including models that involve Born theory to estimate bulk hydration 
contributions for large ions \cite{deligny}. Those estimated bulk contributions are subtracted from the so-called {\it real} hydration 
free energies (the free energy terminology is discussed below), allowing estimation of a surface potential contribution ($-0.3$ V in Ref.~\cite{deligny}).  Alternatively, Randles 
discusses studies of the temperature dependence of cell emfs \cite{randles77}; early experiments suggested a small positive surface 
potential value (0.1 to 0.2 V) \cite{randles65}, but the final result quoted in Ref.~\cite{randles77} is $0.08 \pm 0.06$ V, or effectively zero.  Subtleties related to the effects of interfacial potentials on single-ion thermodynamic properties are discussed in the recent books of Fawcett \cite{fawcett-book} and H\"{u}nenberger and Reif \cite{hunenberger-book}.

How can the widely differing electron holography and electrochemical results be related?  Kathmann {\it et al.}~\cite{kathmann_understanding_2011} have taken an important step toward answering this question by showing that, with omission of certain regions of space (namely the space inside the water molecules) in computing average potentials, the computed and electrochemically estimated surface potentials come into closer agreement 
(as do the classical and quantum mechanically calculated results). 
In related work involving classical point-charge and polarizable simulations, Harder and Roux \cite{harder_origin_2008} and Vorobyov and Allen \cite{allen_SP} have shown that it is the real hydration free energy, as opposed to the {\it intrinsic} free energy (also defined below),  that provides relatively consistent results due to a cancellation of model-dependent molecular quadrupole contributions between the solute-solvent and water liquid-vapor boundaries.  Two goals of the present paper are to 
extend the analysis in Refs.~\cite{harder_origin_2008} and \cite{allen_SP} to quantum calculations and to provide a firmer statistical mechanical resolution of the
above question.  In the process, the surface potential as defined by Landau, Lifshitz, and Pitaevskii \cite{landau8}, and that inferred from electrochemical experiments \cite{randles77}, are shown to be different quantities.   
 
The discussion presented here impacts on both measurements of single-ion absolute hydration free energies and the distributions of ions near the water surface.  It will be shown that the proton (real) hydration free energy estimate obtained by combining bulk and cluster data, and using the cluster-pair approximation \cite{mtiss98}, does include a full contribution from the bulk water surface potential under the assumption that the common intersection behavior observed in the data analysis continues to hold for clusters larger than $n = 6$.  Differences between the proton hydration free energy reported in Ref.~\cite{mtiss98} and other tabulations involving extra-thermodynamic assumptions \cite{marcus,rschm00} are largely due to omission of the electrochemical surface potential, which is shown below to be closely related to the net potential at the center of a neutral, ion-sized particle.  For typical monatomic ions, that net potential shows limited size dependence \cite{ghumm96,hank-phi,garde-phi,droge10}, and persists up to relatively large solute sizes \cite{hank-phi}. 

In previous classical studies, it has been shown that the net potential influences the distributions of ions near the water surface \cite{tlbsurf,mundy-SP-12}. But the question remains whether this influence persists when water is modeled at the quantum mechanical level \cite{mundy-SP-12}.  The statistical mechanical basis for addressing this question is first developed, and results are then presented for an ion in a water droplet (of sizes $n=105$ and $n=242$). Again, the net potential at the center of a neutral particle is shown to play a prominent role. The quantum calculations are performed on clusters at the all-electron DFT and Hartree-Fock levels, thus eliminating potential complications arising from pseudo-potentials and  intricacies of the Ewald potential in periodic boundaries.  The classical droplet results (for $n=105$ and $n=511$ clusters) are similar to the previous classical simulations in the slab geometry \cite{tlbsurf}. The quantum results for the net potential are qualitatively similar to the classical point-charge (SPC/E water model) results. As in the classical studies discussed above, the reason for the similarity of the net potentials (in spite of the enormous difference between the computed water surface potentials) is due to the cancellation of a quadrupole contribution that differs widely (in sign and magnitude) between point-charge and quantum water models.  In the quantum case the canceling terms are significantly larger. 

\section{Outline}

The main focus in this paper will be on the simple situation of a single ion inside a water droplet.  The bulk (infinite dilution) limit is attained as $n \rightarrow \infty$, where $n$ is the number of water molecules. To make contact with previous studies \cite{tlbsurf} and recent experiments \cite{saykally12} (on a related large, chaotropic ion, SCN$^-$), the I$^-$ ion is primarily examined in the calculations. The hydration free energy of the Na$^+$ ion is also computed.  It will be assumed that the ions are monatomic and monovalent to ease the notation. 

The discussion begins with a review of the electrostatics of the surface potential and its decomposition into molecular dipole and quadrupole contributions.  Then two forms for the ion hydration free energy are discussed that will be useful in analyzing surface potential effects.  Next, statistical mechanical expressions are derived for the hydration free energy, and connection is made to a simple and robust approach for extracting bulk single-ion quantities from ion hydration data.  Two exact approaches for partitioning the free energy are summarized; the partitioning aids both in relating to experimental measurements and in isolating the role of the surface potential in the free energy. 

The discussion then turns to recent experimental determinations of the absolute hydration free energy of the proton \cite{mtiss98}, and adsorption enthalpy and entropy changes for a large anion approaching the water liquid-vapor surface \cite{saykally12}.  At first glance, these experiments would appear to be unrelated. It is shown, however, that the same free energy shift included in the proton hydration free energy measurement leads to a net driving force that enhances anion density relative to cations at the water surface.  Results of classical and quantum calculations on ions in water droplets are presented that provide preliminary support to this view.  

\section{The surface potential}

The potential through a liquid-vapor interface (starting in the vapor, $v$, and passing to the liquid, $l$) can be obtained by integrating the one-dimensional Poisson equation \cite{pratt_contact_1992}: 
\begin{equation}
\phi (z)  - \phi (v) = 4\pi \int_{v}^{z}  (z' - z) \rho_q (z') dz'
\label{eq:surfpotl}
\end{equation}
where $z$ is directed along the surface normal, and $\rho_q (z)$ is the total average charge density at a particular $z$ location.
In the macroscopic limit, integrating deep into the liquid, the resulting surface potential is \cite{landau8}
\begin{equation}
\phi_{sp} = 4\pi \int_{v}^{l}  z \rho_q (z) dz
\label{eq:surfpotl1}
\end{equation}
Thus, the surface potential jump is the result of a dipole layer. That surface dipole in turn results from the average distribution of all the charges, however, and does not simply reflect molecular dipole orientations at the interface.  In simulations, it has been shown that the potential shift across the water surface occurs on a length scale of less than 10~{\AA} \cite{warren_hydration_2007,wick_effect_2007}. 

In pioneering work, Wilson, Pohorille, and Pratt \cite{mwils88,lrp_SP_89} showed that the surface potential of water computed from the molecular dipole contributions alone yields incorrect results.  Building on the development of Jackson \cite{jackson} for the average macroscopic charge distribution in terms of the molecular multipole contributions, they showed that there is also a substantial contribution to the surface potential from the molecular quadrupoles:
\begin{equation}
\phi_{sp} = 4\pi \int_{v}^{l} P_z (z') dz' 
  - 4 \pi \left[ Q_{zz} (z_l) - Q_{zz} (z_v) \right]
\label{eq:surfpotllrp}
\end{equation}
where $P_z$ is the $z$ component of the molecular dipole density, and $Q_{zz}$ is the $zz$ component of the density of molecular quadrupoles.  The separate dipole and quadrupole contributions depend on the choice of molecular center, but their sum does not.  The multipole expansion truncates sharply at the quadrupole term since it is a gradient expansion, and the two integrations leave spatial derivatives for the higher-order terms that are zero in the bulk phases. 

The density of quadrupoles in the vapor is negligible, and the quadrupole density in the liquid is given by the trace of the quadrupole tensor:
\begin{equation}
Q_{zz} (z_l) = \frac{1}{3} \mathrm{Tr} \left[ Q_{\alpha \gamma} (z_l) \right] =
\left\langle \sum_m \delta (z_l - z_m) \left( \frac{1}{2}\sum_i q_{im} z_{im}^2 \right) \right\rangle
\label{eq:quadrupole}
\end{equation}
for a molecular point-charge model \cite{warren_hydration_2007}. The $m$ label is for the molecules, while $i$ labels the charges within a molecule. 
For the quantum mechanical case, the charges are replaced by the continuous charge density, and the sum over $i$ is replaced by an integral; the general formula for the quadrupole matrix elements is then \cite{jackson}:
\begin{equation}
Q_{\alpha \gamma} = 3 \int x_{\alpha}' x_{\gamma}' \rho ({\bf x}') d^3 x' 
\label{eq:quadrupole1}
\end{equation}
where $x_{\alpha}$ are the Cartesian $x, y, z$ components and $\rho ({\bf x})$ is the charge density.  Jackson \cite{jackson} provides 
a careful discussion of why the nonzero-trace form of the quadrupole is required for the expansion of the charge density, as 
opposed to the traceless form (the trace adds a required $l=0$ term to the multipole expansion).  

An interesting feature of these results is that the quadrupole contribution is a property of the bulk liquid and has nothing to do with molecular orientations at an interface.  We will see below that the classical point-charge quadrupole trace for water is positive, leading to a negative contribution to the surface potential, while the quantum result is large and negative, producing a large positive contribution.  Another point is that, if we consider the net potential inside an ion-sized cavity embedded deep in a water droplet, the quadrupole contribution cancels since the quadrupole densities both in the vapor and inside the cavity are zero.  The process of moving a non-interacting test charge from the vapor, into the droplet, and then into the cavity results in the crossing of two `interfaces', the water liquid-vapor interface, and the water-cavity boundary, leading to the cancellation.  Then the net potential at a cavity center reflects a dipole contribution. But that dipole contribution arises from differences in water orientations between the solute-water boundary and the water liquid-vapor interface.  

\section{Ion hydration free energy definitions}

The above discussion views the surface potential as an electrostatic effect involving the average charge distribution through the interface.  The equilibrium electrochemical potential $\mu$ for a given ion is typically written as \cite{guggenheim,schmickler,hamann} 
\begin{equation}
\mu =  kT \ln \left[ \rho \Lambda^3 \right] + \mu^{ex} = kT \ln \left[ \rho \Lambda^3 \right] + \mu_{int}^{ex} + q\phi_{sp}
\label{eq:gibbs}
\end{equation}
where $k$ is Boltzmann's constant, $T$ is the temperature, $\rho$ is the ion number density,  $\Lambda$ is the thermal de Broglie wavelength, $\mu^{ex}$ is the real hydration free energy, $\mu_{int}^{ex}$ is the intrinsic hydration free energy (in the absence of an interface) \cite{lamoureux_absolute_2006}, and $q$ is the ion charge. The real hydration free energy $\mu^{ex}$ is the free energy change for moving an ion from a vacuum, across the water surface, and into the bulk. The surface potential is the difference of the inner (Galvani) and outer (Volta) potentials \cite{schmickler}.
In what follows we will assume that the potential is zeroed in one of the phases, here taken as the vapor phase. Both $\mu^{ex}$ and $\phi_{sp}$ can be obtained experimentally ($\mu^{ex} = -59.3$ kcal/mol for I$^-$ \cite{mtiss98} while $q\phi_{sp} = -80.7$ kcal/mol estimated from electron holography \cite{holography}), resulting in $\mu_{int}^{ex} = +21.4$ kcal/mol.  These results show that $\mu_{int}^{ex}$ differs from any electrochemical estimates of bulk hydration free energies \cite{leung_ab_2009,leung_surface_2010}.  (The DFT calculations of ion hydration free energies in Refs.~\cite{leung_ab_2009,leung_ab_2009_rep} included a procedure to subtract off the large quadrupole contribution arising in periodic boundaries; that procedure leads to free energies more in line with the bulk values discussed below.) 

To set the stage for the discussion below, and following Harder and Roux \cite{harder_origin_2008}, we choose to re-define the real hydration free energy as
\begin{equation}
\mu^{ex} = \mu_{b}^{ex} + q\phi_{np}
\label{eq:gibbs1}
\end{equation}
where $\mu_{b}^{ex}$ is a bulk hydration free energy, while $\phi_{np}$ is the net potential at the center of an uncharged solute.  The rearrangement amounts to adding and subtracting a term which is the charge times the local potential due to nearby waters sampled with an uncharged solute.  As discussed above, this leads to a cancellation of quadrupole terms.  With the results obtained below, the resulting I$^-$ $\mu_{b}^{ex}$ value is $-70.9$ kcal/mol (using the value $\phi_{np} = -11.6$ kcal/mol-e or $-0.50$ V).  With this re-definition, 
$\mu_{b}^{ex}$ is much closer to estimates based on the parameterization of Latimer, Pitzer, and Slansky \cite{latimer_39} ($\mu_{b}^{ex} = -71.4$ kcal/mol in Ref.~\cite{ashbaugh_single_2008}, see below). We note here that all hydration free energies listed in the present paper refer to a transition from a 1 M vapor to a 1 M solution phase concentration.  Thus, for example, all free energies taken from Refs.~\cite{mtiss98} and \cite{ashbaugh_single_2008} are uniformly shifted downward by 1.9 kcal/mol. 

It is clear that surface potential estimates indirectly obtained from electrochemical measurements (and perhaps further assumptions) are close in magnitude to the $\phi_{np}$ value listed above ($-0.50$ V).  The cost of the rearrangement, however, is a slight solute-size specificity to $\phi_{np}$ (of magnitude 1 kcal/mol-e variation for neutral particles sampled with Lennard-Jones potentials) \cite{ghumm96,garde-phi,droge10}.  

To summarize, six quantities have been introduced: the electrochemical potential $\mu$, the real hydration free energy $\mu^{ex}$, the intrinsic hydration free energy $\mu_{int}^{ex}$, the bulk hydration free energy $\mu_{b}^{ex}$, the surface potential $\phi_{sp}$, and the net potential at the center of an uncharged solute $\phi_{np}$.  The terminology used here was chosen to follow the existing electrochemical and simulation literature \cite{schmickler,lamoureux_absolute_2006}. All of the excess quantities and the potentials can be estimated from experiment, but the surface potential cannot be measured thermodynamically.  

\section{Statistical mechanics of ion hydration}

We take the Potential Distribution Theorem (PDT) as the fundamental expression for the chemical potential  \cite{bwido63,ourbook}:
\begin{equation}
\mu = kT \ln \left[ \rho \Lambda^3 \right] + \mu^{ex}
\label{eq:mu}
\end{equation}
where
\begin{equation}
\mu^{ex} = -kT \ln \left< \exp (-\varepsilon / kT) \right>_0
\label{eq:pdt}
\end{equation}
and $\varepsilon$ is the interaction energy of the ion with the surrounding medium.  The `0' subscript on the thermal average implies the sampling is conducted with no ion present.  Eq.~\ref{eq:pdt} is equivalent to the inverse form
\begin{equation}
\mu^{ex} = kT \ln \left< \exp (\varepsilon / kT) \right>
\label{eq:invpdt}
\end{equation}
where now the sampling includes the ion.  The excess chemical potential in Eqs.~\ref{eq:pdt} and \ref{eq:invpdt} includes any possible contribution from the surface potential since that contribution arises from the interaction energy of the ion with a distant dipole layer. 

Above, the uncoupled sampling in Eq.~\ref{eq:pdt} has no ion present; this is a correct formula for the excess chemical potential, but can lead to computational difficulties for larger particles due to the hard ion core.  Thus in what follows we will assume that the free energy has already been estimated for the uncharged ion, and the `0' subscript will imply sampling with the uncharged ion present but no electrostatic coupling. The `{\it es}' subscript indicates the electrostatic contribution to the energy or free energy. Then the total hydration free energy is $\mu^{ex} = \mu_{vdW}^{ex} + \mu_{es}^{ex}$, or the sum of the non-polar and electrostatic contributions. 

The resulting electrostatic contributions to Eqs.~\ref{eq:pdt} and \ref{eq:invpdt} can be rewritten exactly as
 \begin{equation}
\mu_{es}^{ex} = \left< \varepsilon_{es} \right>_0 - kT \ln \left< \exp \left[ -(\varepsilon_{es} - \left< \varepsilon_{es} \right>_0) \right] / kT) \right>_0
\label{eq:pdt1}
\end{equation}
and
\begin{equation}
\mu_{es}^{ex} = \left< \varepsilon_{es} \right> + kT \ln \left< \exp \left[ (\varepsilon_{es} - \left< \varepsilon_{es} \right>) \right] / kT) \right>
\label{eq:invpdt1}
\end{equation}
Eqs.~\ref{eq:pdt1} and \ref{eq:invpdt1} express the excess chemical potential as the sum of an average binding energy plus a fluctuation contribution involving deviations from the mean.  

Taking the average of Eqs.~\ref{eq:pdt1} and \ref{eq:invpdt1} and rearranging, we obtain
\begin{equation}
\mu_{es}^{ex} = \left\langle \varepsilon_{es} \right\rangle_0  +
\frac{1}{2} \left( \left\langle \varepsilon_{es} \right\rangle - \left\langle \varepsilon_{es} \right\rangle_0
\right) + \frac{kT}{2} \left( \ln \left\langle \exp [ ( \varepsilon_{es} - <\varepsilon_{es}> )/kT ] \right\rangle  -
 \ln \left\langle \exp [ -( \varepsilon_{es} - <\varepsilon_{es}>_0 )/kT ] \right\rangle_0 \right)
\label{eq:pdtnp}
\end{equation}
The first term on the rhs of Eq.~\ref{eq:pdtnp} is $q\phi_{np}$, where $\phi_{np}$ is the average net potential at the center of an uncharged solute.  Then we can associate the bulk hydration free energy, $\mu_b^{ex}$ above, with the sum of the non-polar contribution $\mu_{vdW}^{ex}$ and all of the terms on the rhs of Eq.~\ref{eq:pdtnp} except the first.  The second term above involves the difference of the electrostatic interaction energy sampled with full coupling and with no electrostatic interactions; it has a substantial magnitude but contains no contribution from the surface potential.  This term will contain a local contribution that is ion specific and a far-field contribution that does not depend on the ion charge or size and could be well-handled with continuum models.  The final term on the rhs involves an average of fluctuation contributions.  Since previous work has shown that free energy shifts due to the surface potential can be handled at the mean-field level \cite{tlbsurf}, this term thus includes little or no contribution from the surface potential. Further, it exhibits ion specificity due to relatively local interactions \cite{tlbent11}. If the interaction energy distributions were exactly Gaussian, then the average of the fluctuation terms would be zero (below).

Considering now the first $q\phi_{np}$ contribution to the free energy, the net potential  $\phi_{np}$ is obtained by sampling with no charge on the particle.  If we assume the particle is situated at the center of a large water droplet, this potential will contain contributions from waters nearby the particle and from any asymmetric charge distribution at the distant water surface.  Since all waters between these two domains will be randomly oriented, they contribute nothing to the net potential.  Thus we can express the net potential as \cite{harder_origin_2008}
\begin{equation}
\phi_{np} = \phi_{lp} + \phi_{sp}
\label{eq:netpotential}
\end{equation}
where $\phi_{lp}$ is the local potential contribution from nearby waters.  Based on the above discussion, both $\phi_{lp}$ and $\phi_{sp}$ contain large but canceling quadrupole contributions, leaving behind a net potential that reflects differences in water dipole orientations between the solute-solvent boundary and the liquid-vapor interface.  

Extensive previous classical point-charge simulation work has shown that, for these models, the value of the local potential $\phi_{lp}$ is positive and of magnitude roughly $8-9$ kcal/mol-e \cite{ghumm96,hank-phi,garde-phi}.  The previous simulations have been performed in periodic boundaries, so the average potential at the center of the uncharged particle is exactly $\phi_{lp}$, with no $\phi_{sp}$ contribution from a distant interface.  Addition of the $\phi_{sp}$ term ($-13.8$ kcal/mol-e for the SPC/E model) results in the $\phi_{np}$ value of $-5.7$ kcal/mol-e for the neutral iodide particle in Ref.~\cite{tlbsurf}.  For similar uncharged solute particles, Rajamani, Ghosh, and Garde \cite{garde-phi} observed a variation in $\phi_{lp}$ of about 1 kcal/mol-e for modest-sized particles (all but the smallest Li$^+$ ion), displaying the small ion specificity to this quantity.  Interestingly, Ref.~\cite{hank-phi} shows that negative $\phi_{np}$ values persist up to large particle sizes of magnitude 15~{\AA} (since the observed magnitude of $\phi_{lp}$ is smaller
than that for $\phi_{sp}$), and the $\phi_{np}$ values are relatively size-independent beyond 5~{\AA}.  At very large sizes, $\phi_{np}$ should converge to zero for hard core solutes (as employed in Ref.~\cite{hank-phi}), but this convergence does not occur for molecular-sized solutes.  

Performing a second-order cumulant expansion of Eq.~\ref{eq:pdtnp}, we obtain
\begin{equation}
\mu_{es}^{ex}\approx \left\langle \varepsilon_{es} \right\rangle_0  +
\frac{1}{2} \left( \left\langle \varepsilon_{es} \right\rangle - \left\langle \varepsilon_{es} \right\rangle_0
\right) + \frac{1}{4kT} \left( \left\langle \delta \varepsilon_{es}^2 \right\rangle -
\left\langle \delta \varepsilon_{es}^2 \right\rangle_0 \right)
\label{eq:pdtnp1}
\end{equation}
where $\delta \varepsilon_{es}$ is the deviation from the mean. Ref.~\cite{ghumm96} shows that the difference of the fluctuation terms is very small for cations but of sizable magnitude for anions (with $\left\langle \delta \varepsilon_{es}^2 \right\rangle > \left\langle \delta \varepsilon_{es}^2 \right\rangle_0$).   Outside a transition region centered at a small negative charge, Ref.~\cite{ghumm96} also suggests that linear response theory is reasonably accurate separately for cations and anions, but with different slopes. The cation/anion difference is illustrated by the result in Ref.~\cite{tlbloc11} that a simple Gaussian estimate of the free energy is modestly accurate for both electrostatically coupled and uncoupled sampling states for cations, but is only sensible for the coupled sampling state for anions.  These differences arise from the penetration of the water protons closer to the anions.  

Alternatively, the second-order expansion of the electrostatic part of Eq.~\ref{eq:pdt} yields
\begin{equation}
\mu_{es}^{ex} \approx \left< \varepsilon_{es} \right>_0 - \frac{1}{2kT} \left< \delta \varepsilon_{es}^2 \right>_0 = q \phi_{np} - \frac{q^2}{2kT} \left< \delta \phi_{np}^2 \right>_0 \approx
q \phi_{np}  - \frac{q^2}{2R} \left( 1 - \frac{1}{\epsilon} \right)
\label{eq:pdtgauss}
\end{equation}
Eq.~\ref{eq:pdtgauss} displays the equivalence to a Born model with the addition of the mean-field term  
$q\phi_{np}$. The Born radius is related to the potential fluctuations at the uncharged solute center, and is thus temperature dependent.  
It has been shown previously that the distribution of total electrostatic interaction energies is not accurately Gaussian, and the deviations from Gaussian behavior are due to relatively local interactions between the ion and nearby waters \cite{tlbloc11}.  

\section{Estimating $\phi_{np}$}

Starting from the model of Latimer, Pitzer, and Slansky (LPS) \cite{latimer_39}, Ashbaugh and Asthagiri \cite{ashbaugh_single_2008} showed that, by shifting cation and anion radii by a constant distance (one distance for cations and one for anions),  and adding one other parameter that mimics the free energy for inserting the neutral particle, 
\begin{equation}
\mu^{ex} (\mathrm{pair} ) = -\frac{q^2}{2} \left( 1 - \frac{1}{\epsilon} \right) \left( \frac{1}{r_+ + \delta_+} + \frac{1}{r_- + \delta_-} \right) + 2 \lambda 
\label{eq:lps}
\end{equation}
an extremely good linear fit is observed between Born and experimental hydration free energies for a wide range of monatomic ion pairs.  Fitting to the pair free energies largely cancels  any contribution from the net potential (with the caveat of the slight size-dependence to $\phi_{sp}$).  These results suggest that the observed deviation from Gaussian behavior can be handled with two radius shift parameters.  The smaller computed radius shift for anions is due to the observed stronger hydration \cite{ghumm96} of anions compared with cations.  

After shifting all the single-ion free energies in Ref.~\cite{ashbaugh_single_2008} due to the standard state correction of -1.9 kcal/mol discussed above, the cation and anion data in the second column of their Table III are shifted by an average magnitude of 11.6 kcal/mol from experiment \cite{mtiss98} (with small deviations of order 0.5 kcal/mol). The shifts are quite uniform along the entire cation and anion series.  Here it is suggested that this shift implies a $\phi_{np}$ value of -11.6 kcal/mol-e ($-0.5$ V), and we consider this an experimental value.  The fact that this result for the net potential $\phi_{np}$ is so close to the calculated value of $\phi_{sp}$ for point-charge classical models \cite{tildesley_wsp,warren_hydration_2007,tlbsurf} has led to some confusion in the literature regarding these two quantities ($\phi_{sp}$ and $\phi_{np}$); the quantum mechanical results presented below highlight the large difference between them. 

Ref.~\cite{ashbaugh_single_2008} also noted that free energies calculated from simulation (in periodic boundaries) are shifted from the LPS estimates in one direction for cations and the other for cations; these shifts are due to $\phi_{lp}$ (due to simulating in periodic boundaries), and not to $\phi_{np}$.  Also, when the $\phi_{lp}$ shift is taken into account, the predicted bulk  hydration free energies (here closely related to $\mu_{b}^{ex}$ above) are nearly identical for large cations and anions. This result shows that previous simple models used to predict the electrochemical surface potential, or $\phi_{np}$, by subtracting a Born bulk estimate from the real hydration free energy of a large cation (with a result of $-0.3$ V) hold some validity \cite{deligny}. 

\section{Partitioning the free energy}

Dividing up free energy contributions due to different regions of space, or different interactions, has a long history.  In recent years, two such divisions \cite{lrprat98,lrprat991,ourbook,lrp_coordination_10,rodgers_local_2008} have been derived that start from exact statistical mechanical theory.  An exact formulation is helpful, since then the accuracy of subsequent approximations can be assessed.  

The Quasichemical Theory (QCT) \cite{ourbook} involves a spatial decomposition of the free energy into inner-shell and outer-shell components.  This partitioning is effected by manipulations involving repulsive particles.  Here we will call the potential of interaction between the repulsive particle and the solvent $M$ (model potential); the original QCT utilized a hard sphere particle for $M$, but a `soft-cutoff' version has been derived \cite{chempath_quasichemical_2009} that allows for analysis using repulsive particles with continuous potentials.  As in Ref.~\cite{dilip-regularize}, the PDT can be re-written as 
\begin{equation}
\mu^{ex} = kT \ln \left\langle \exp (-M/kT) \right\rangle  - kT \ln \left\langle \exp (-M/kT) \right\rangle_0 - kT \ln \left\langle \exp (-\varepsilon/kT) \right\rangle_M 
\label{eq:qct}
\end{equation}
where $\varepsilon$ is the full solute-solvent interaction energy, and the sampling in the last term is conducted with the repulsive $M$-particle included.  The first term on the right is an inner-shell contribution, the second term is a packing contribution that is the free energy to insert the $M$-particle, and the last term is the outer-shell, long-ranged contribution that involves solute-solvent interactions with the solvent pushed away from the solvent a distance specified by $M$.  Physically, the inner-shell term reflects local (chemical) interactions of the solute with the solvent, and is minus the work required to push the nearby solvent out to a length scale specified by the $M$ potential. The packing term is the free energy of cavity formation, and the long-ranged term includes all other interactions {\it conditioned} on the lack of solvent molecules in the inner shell. 

More recently, the Local Molecular Field Theory (LMFT) approach of Weeks and coworkers \cite{rodgers_local_2008,rodgers_interplay_2008,rodgers_accurate_2009} has been adapted to calculations of free energies \cite{tlbloc11}.  In terms of electrostatic interactions, the LMFT approach is to partition the interactions following the Ewald prescription:
\begin{equation}
\frac{1}{r} = \frac{\mathrm{erfc} (\eta r) }{r}  + \frac{\mathrm{erf} (\eta r)}{r}
\label{eq:ewald}
\end{equation} 
where $\eta^{-1}$ specifies the length scale for the partitioning.  In our studies of ion hydration \cite{tlbloc11,tlbent11,tlbsurf}, that length scale has been chosen to mainly involve the first hydration shell, so $\eta^{-1} \approx 4-5$~{\AA}.  The first term yields a local electrostatic contribution, while the second includes all distant (far-field) interactions.  In the Ewald method, the second term is transformed to a ${\bf k}$-space representation.  

Using this LMFT partitioning, and first partitioning out the van der Waals (vdW) contribution, the free energy can then be written exactly as 
\begin{equation}
\mu^{ex} = \mu_{vdW}^{ex} + kT \ln \left\langle \exp (\varepsilon_{es,loc}/kT) \right\rangle_{loc}
+ kT \ln \left\langle \exp (\varepsilon_{es,far}/kT) \right\rangle
\label{eq:lmft}
\end{equation}
Note that the sampling in the two electrostatic terms is conducted on different potentials, the local potential for the local term, and the fully coupled potential for the far-field term.  Eq.~\ref{eq:lmft} is equivalent to 
\begin{equation}
\mu^{ex} = \mu_{vdW}^{ex} + kT \ln \left\langle \exp (-\varepsilon_{es,loc}/kT) \right\rangle_{vdW}
+ kT \ln \left\langle \exp (-\varepsilon_{es,far}/kT) \right\rangle_{loc}
\label{eq:lmft1}
\end{equation}
It was found in our models of ion hydration in bulk water (periodic boundaries, free of interfaces) and in the slab geometry \cite{tlbloc11,tlbsurf}, that the far-field electrostatic contribution is Gaussian distributed to high accuracy.    

Using the above LMFT partitioning, Ref.~\cite{tlbsurf} showed that the free energy shift for an ion moving from a bulk periodic boundaries situation to the slab geometry is exactly the same as $q\phi_{sp}$, where $\phi_{sp}$ is determined from the integral of the charge density in Eq.~\ref{eq:surfpotl1}.  This is a direct confirmation that Eq.~\ref{eq:gibbs} is appropriate, albeit with large canceling contributions from the intrinsic free energy and the surface potential.  This approach of adding a surface potential contribution to the intrinsic hydration free energy obtained in periodic boundary simulations has been used widely recently in comparing single-ion (real) free energies to experiment during force-field development \cite{lamoureux_absolute_2006,warren_hydration_2007,horinek_rational_2009}. 

The above discussion of energetic partitioning also provides a direct indication of why $\phi_{sp}$ is not measurable thermodynamically: while $\mu^{ex}$ can be measured \cite{mtiss98}, separation into the intrinsic hydration free energy and surface potential parts involves an extra-thermodynamic partitioning of the electrostatic portion of the free energy. This point was noted by Guggenheim \cite{guggenheim} and Randles and Schiffrin \cite{randles65}. 

\section{The cluster-pair approximation and the hydration free energy of the proton}

In 1998, Tissandier {\it et al.} experimentally determined the absolute hydration free energy of the proton \cite{mtiss98}.  They reported the value $\mu_{\mathrm{H}^+}^{ex} = -265.9$~kcal/mol, obtained by combining cluster hydration data with bulk hydration data in a clever way (the cluster-pair approximation, CPA).  Once an absolute hydration free energy is obtained for one ion, single-ion free energies for all other ions can be obtained from accurate thermodynamic data.  

Here we briefly summarize the CPA in order to analyze whether the surface potential appears in the experimental results, and if so, where.  We use the recently derived elegant and exact QCT formula from Asthagiri {\it et al.}  \cite{lrp_coordination_10}
\begin{equation}
\mu^{ex} = - kT \ln K_n^{(0)} \rho_W^n + kT \ln p (n) + \mu_{XW_n}^{ex} - n \mu_W^{ex}
\label{eq:qctexact}
\end{equation}
where $K_n^{(0)}$ is the equilibrium constant for binding $n$ waters to the ion $X$ {\it in the gas phase}, $\rho_W$ is the bulk density of water, $p (n)$ is the probability of observing $n$ waters in the inner-shell specified by a chosen cutoff radius, $\mu_{XW_n}^{ex}$ is the hydration free energy of the $XW_n$ cluster, and $\mu_W^{ex}$ is the hydration free energy of water in water.  This formula is correct for any chosen $n$ value, so long as that $n$ does not exceed a limit imposed by the size of the observation volume.  A sensible choice for $n$ would be the value closest to the mean number of waters in the chosen observation volume, $\bar{n}$. Then making the approximation $p (n) \approx 1$ will not lead to serious errors in the free energy.  If that approximation is made, then the previously derived `primitive' QCT formula is obtained \cite{asthagiri_quasi-chemical_2003}.  

We collect the first, second, and last terms together to obtain an effective free energy for inserting the ion into an $n$-water cluster:
\begin{equation}
\mu_{n}^{ex} = - kT \ln \left[ K_n^{(0)} \rho_W^n/ p (n) \right] - n \mu_W^{ex}
\label{eq:qctcpa}
\end{equation}
leading to
\begin{equation}
\mu^{ex} = \mu_{n}^{ex} + \mu_{XW_n}^{ex} 
\label{eq:qctexact1}
\end{equation}
The formula allows us to connect exact statistical mechanics to the CPA.  

The CPA starts by defining conventional hydration free energies as deviations from proton
values.  Here we consider monovalent ions and label cations as $P$ and anions as $N$. Then
\begin{equation}
\mu_P^{ex,con} = \mu_P^{ex} - \mu_{\mathrm{H}^+}^{ex}
\label{eq:exconp}
\end{equation}
and 
\begin{equation}
\mu_N^{ex,con} = \mu_N^{ex} + \mu_{\mathrm{H}^+}^{ex}
\label{eq:exconn}
\end{equation}
These conventional free energies are available from thermodynamically measured free energy tabulations. An important point to note is that the conventional ion hydration free energies then include no contribution from the surface potential.

Similarly, the conventional free energies for hydrating the clusters can be defined as 
\begin{equation}
\mu_{PW_n}^{ex,con} = \mu_{PW_n}^{ex} - \mu_{\mathrm{H}^+}^{ex}
\label{eq:exconpclus}
\end{equation}
and 
\begin{equation}
\mu_{NW_n}^{ex,con} = \mu_{NW_n}^{ex} + \mu_{\mathrm{H}^+}^{ex}
\label{eq:exconnclus}
\end{equation}

Using the above results, the following formula can be derived:
\begin{equation}
\frac{1}{2} \left[ \mu_{NW_n}^{ex,con} - \mu_{PW_n}^{ex,con} \right] = 
\frac{1}{2} \left[ \mu_{NW_n}^{ex} - \mu_{PW_n}^{ex} \right] +
\mu_{\mathrm{H}^+}^{ex}
\label{eq:cpa1}
\end{equation}
which is the same as 
\begin{equation}
\frac{1}{2} \left[ 
\mu_{N}^{ex,con} - \mu_{P}^{ex,con} - 
\left( \mu_{N,n}^{ex} - \mu_{P,n}^{ex} \right)
\right] 
= 
\frac{1}{2} \left[ \mu_{NW_n}^{ex} - \mu_{PW_n}^{ex} \right] +
\mu_{\mathrm{H}^+}^{ex}
\label{eq:cpa2}
\end{equation}
The quantities on the left can be measured thermodynamically and in cluster experiments.  The first term on the right clearly goes to 0 as the number of waters increases.  Following the discussion of Kelly, Cramer, and Truhlar \cite{kelly_aqueous_2006}, data for a wide range of ions can be plotted in the form: $x$-axis as $[ \mu_{N}^{ex,con} - \mu_{P}^{ex,con}]/2$ and $y$-axis as $[ \mu_{NW_n}^{ex,con} - \mu_{PW_n}^{ex,con}]/2$. For cluster studies (involving multiple ions) with sizes up to $n=6$ water molecules, what is observed is that these curves all cross at a common point on the $y$-axis (see Fig.~1 in Ref.~\cite{kelly_aqueous_2006}).  The slope of the curves decreases in magnitude with increasing $n$: for large $n$ the curves should be flat at the value of the proton hydration free energy, since the first term on the rhs approaches 0 for large $n$. Assuming that the crossing point location holds as $n$ increases, then the full lhs value is a constant for all $n$ at the crossing point. This suggests that the value of the first term on the rhs, {\it for this hypothetical ion pair}, is 0 for all $n$, leading to an accurate free energy estimate even for small cluster sizes.  The result of this analysis is not simply a cluster free energy; rather the cluster data is used to tease out an estimate of the bulk single-ion value. The only `approximation' appears to be the assumption of continuation of the crossing behavior up to larger $n$. 

Given the above discussion, we can ask whether the surface potential is included in the reported free energy of the proton.  Asthagiri, Pratt, and Ashbaugh \cite{asthagiri_absolute_2003} have previously concluded that $\phi_{sp}$ is indeed included in the reported value of $-265.9$ kcal/mol \cite{mtiss98}.  As discussed above, the conventional ion hydration free energy values contain no contribution from the surface potential.  Then we can see from the left side of Eq.~\ref{eq:cpa2} that, in the limit of large $n$, the surface potential contribution is carried in the difference of the {\it cluster} free energies ($\mu_{N,n}^{ex} - \mu_{P,n}^{ex}$).  Below, ion hydration free energies in clusters with $n=105$ and $n = 511$ waters will be examined. These clusters display surface potential values close to the result computed in the slab geometry; the computed hydration free energy of an I$^-$ ion agrees quite well with the value reported by Tissandier {\it et al.} \cite{mtiss98}.  Thus the results suggest both that the experimental proton value includes a contribution from the bulk water surface potential and that the CPA assumption holds for larger clusters.

\section{Measured enthalpy and entropy changes for anions near the interface}

From the basic PDT forms of the excess chemical potential, Eqs.~\ref{eq:pdt} and \ref{eq:invpdt}, the hydration enthalpy and entropy can be derived from the appropriate temperature derivatives.  Here we will ignore pressure-volume effects. The enthalpy is then
\begin{equation}
h^{ex} = \left< \varepsilon \right> + U_{SR}
\label{eq:enthalpy}
\end{equation}
where
\begin{equation}
U_{SR} = \left< U_{n} \right>_{n + X} - \left< U_{n} \right>_{n} 
\label{eq:sr}
\end{equation}
is the difference in the average energy of the solvent with the ion (labelled $X$) present and with the ion absent (solvent reorganization energy).  The resulting entropy is  
\begin{equation}
s^{ex} = \frac{U_{SR}}{T} - k \ln \left< \exp \left[ (\varepsilon - \left< \varepsilon \right>) \right] / kT) \right>
\label{eq:entropy}
\end{equation}
These results show that the influence of the surface potential from a distant interface is carried predominantly in the enthalpy, not the entropy, since this effect will show up mainly in the mean binding energy $\left< \varepsilon \right>$, with a much smaller contribution from the entropic fluctuation term and no contribution from the solvent reorganization term.  
Ben-Amotz, Raineri, and Stell \cite{ben-amotz-theory1} showed that the fluctuation contribution to the entropy is always negative. As noted above, it has been shown \cite{tlbsurf} that the free energy shift for an ion moving from the bulk to the slab geometries can be accurately modeled with the first-order, mean-field (enthalpic) term. This further supports the notion that the surface potential impact on the entropy is small.  

Also, the development shows that, while the solvent reorganization term is required for estimation of the separate enthalpy and entropy, it cancels out {\it exactly} when compiling the free energy. The cancellation is an example of enthalpy-entropy compensation.  This has been noted previously by Ben-Amotz, Raineri, and Stell \cite{ben-amotz-theory1}. Since the free energy profile determines the driving forces for ions near the interface, care should be taken in associating the term `driving force' with separate enthalpy and entropy contributions. Only those enthalpy and entropy components that can lead to a {\it net} contribution to the free energy can result in a driving force.   It can also be noted that the measured very small magnitudes of the bulk entropies of ion hydration (they are similar to those for the isoelectronic rare gases) result from a near cancellation of a (positive) solvent reorganization term and a (negative) fluctuation term.  

Recent experiments have measured the surface adsorption enthalpy and entropy changes for a large chaotropic anion, SCN$^-$ approaching the water surface from the bulk \cite{saykally12}.   The experiments indicate negative values for both quantities.  The enthalpic term (-2.8 kcal/mol) is larger in magnitude than the entropic contribution $T \Delta s^{ex}$ (-1.2 kcal/mol), leading to a net surface adsorption free energy of -1.6 kcal/mol.  Several simulation studies have explored the enthalpic and entropic contributions \cite{yagasaki_effects_2010,swedes}.  Yagasaki, Saito, and Ohmine \cite{yagasaki_effects_2010} plotted the {\it interaction} parts of Eqs.~\ref{eq:enthalpy} and \ref{eq:entropy} (that is the $U_{SR}$ term was omitted), and observed positive values for the enthalpy and entropy changes.   Caleman {\it et al.} \cite{swedes} and Otten {\it et al.} \cite{saykally12}, on the other hand, computed the full enthalpy and entropy profiles indicated by Eqs.~\ref{eq:enthalpy} and \ref{eq:entropy}, and observed negative values for both quantities, in agreement with experiment.  As indicated above, changes in the water-water interaction term as the ion approaches the surface contribute no net driving force, but lead to the observed negative values for $\Delta h^{ex}$ and $\Delta s^{ex}$. Those separate quantities can yield insights into the ion hydration environment.  It is suggested below that the net potential $\phi_{np}$ provides one part of the net driving force that tends to enhance anion density near the surface relative to cations.

\section{Computational Methods}

The calculations reported here include classical molecular dynamics (MD) simulations and quantum chemical calculations on cluster configurations extracted from the classical simulations.  The classical MD simulations were performed with the Tinker code \cite{tinker}, and the quantum calculations were performed with the ORCA code \cite{ORCA,neese}, both freely available.  

The classical simulations employed the SPC/E water model and used ion parameters from Refs.~\cite{horinek_rational_2009,horinek_specific_2009} (parameter set 2 from Ref.~\cite{horinek_specific_2009}, with $\sigma = 4.09$~{\AA} and $\epsilon = 0.1912$~kcal/mol for the I$^-$ ion and  $\sigma = 2.70$~{\AA} and $\epsilon = 0.1554$~kcal/mol for the Na$^+$ ion).  Clusters of $n = 105$ and $n=511$ water molecules were modeled.  The ion (or neutral particle) was restrained by a strong harmonic force to a particular distance from the center of mass of the water cluster.  An external half-harmonic potential was employed to ensure that any evaporating water molecules were reflected back into the cluster; the starting point for the half-harmonic potential was chosen several {\AA} from the point where the radial density profile decayed to zero, however, so there was little or no interaction of the water molecules with the bounding potential.  A time step of 2 fs was used, and all electrostatic interactions were included.  The clusters were equilibrated for at least 500 ps, followed by production runs of at least 2 ns.

Initial quantum calculations were performed on the $n=105$ water cluster.  Configurations were taken from the MD simulations that employed the SPC/E water model \cite{leung_surface_2010}.  The calculations were performed using Density Functional Theory (DFT) with the BLYP functional and the cc-pVDZ basis set, and included all of the electrons.  This theory/basis combination was chosen to allow for calculations on a large set of 1000 configurations.  The accuracy of the calculations in relation to higher-level methods will be discussed in the results section.  Followup calculations aimed at probing the variation of the net potential with cluster size ($n=242$), theory level (B3LYP and Hartree-Fock), and basis set (Aug-cc-pVDZ) were subsequently performed. The followup calculations were performed on a smaller subset of configurations (typically 109), so the results contain somewhat larger statistical fluctuations in the net potential, and those fluctuations are reported below.

\section{Calculating the surface potential}

As discussed above, the electrostatic profile $\phi(z)$ across the water liquid-vapor interface can be obtained from 
Eq.~\ref{eq:surfpotl}, or equivalently from \cite{wick_effect_2007}
\begin{equation}
\phi(z) - \phi(v) = - \int_{v}^z E_q (z') dz'
\label{eq:dangsurfpotl}
\end{equation}
where $E_q (z)$ is the electric field obtained by integrating the average 
charge density from the vacuum up to $z$. Alternatively, 
$\phi(z)$ can be obtained by simply averaging the Poisson potential computed during {\it ab initio} molecular dynamics (AIMD) simulations \cite{leung_surface_2010,kathmann_understanding_2011}.   While these expressions are exact, averaging the charge density or potential at a particular location results in substantial statistical fluctuations, especially for the quantum case consisting of electrons and the contribution from the pseudo-potential.  

Here, an alternative approach is proposed based on the LMFT (Ewald-like) Coulomb partitioning.  Consider an arbitrary point chosen near the center of mass of a large water cluster.  Then assuming that point is far from the liquid-vapor interface, the average charge density sampled over long times is zero everywhere near the point.  Thus the local charge contribution to the potential at the point is zero.  This conclusion of course holds all the way up to near the interface where the average charge density starts to deviate from zero.   Then the surface potential is simply obtained from the average of the far-field part of the electrostatic potential at the chosen point.  This approach eliminates the large potential fluctuations that occur while averaging the Poisson potential in AIMD simulations.  The far-field contribution to the potential is smooth, and rapid convergence is observed.

Fig.~1 shows the convergence of the surface potential  as a function of simulation time for the $n=511$ SPC/E water cluster.  The surface potential converges to $-14.9$~kcal/mol-e ($-0.65$~V), and is relatively well-converged by  50 ps.  The corresponding surface potential for the $n=105$ cluster is $-14.7$ kcal/mol-e. An advantage of working in the cluster geometry is that there is no ambiguity due to intricacies of the Ewald potential that integrates to zero over the simulation box in periodic boundaries \cite{tlbsurf}.  The surface potential computed for the $n=511$ cluster is more negative than the slab result of Ref.~\cite{tlbsurf} by 1.1 kcal/mol-e (0.05 V).  As shown below, the quantum mechanical result from the far-field potential for the $n=105$ cluster (using the DFT models discussed above) is $+67.0 \pm 0.7$~kcal/mol-e (2.9 V) obtained from only 21 statistically independent configurations. The small standard deviation provides a clear illustration of both the smoothness of the far-field potential and the resulting rapid convergence.

\begin{figure}
\begin{center}
\includegraphics[scale=0.30]{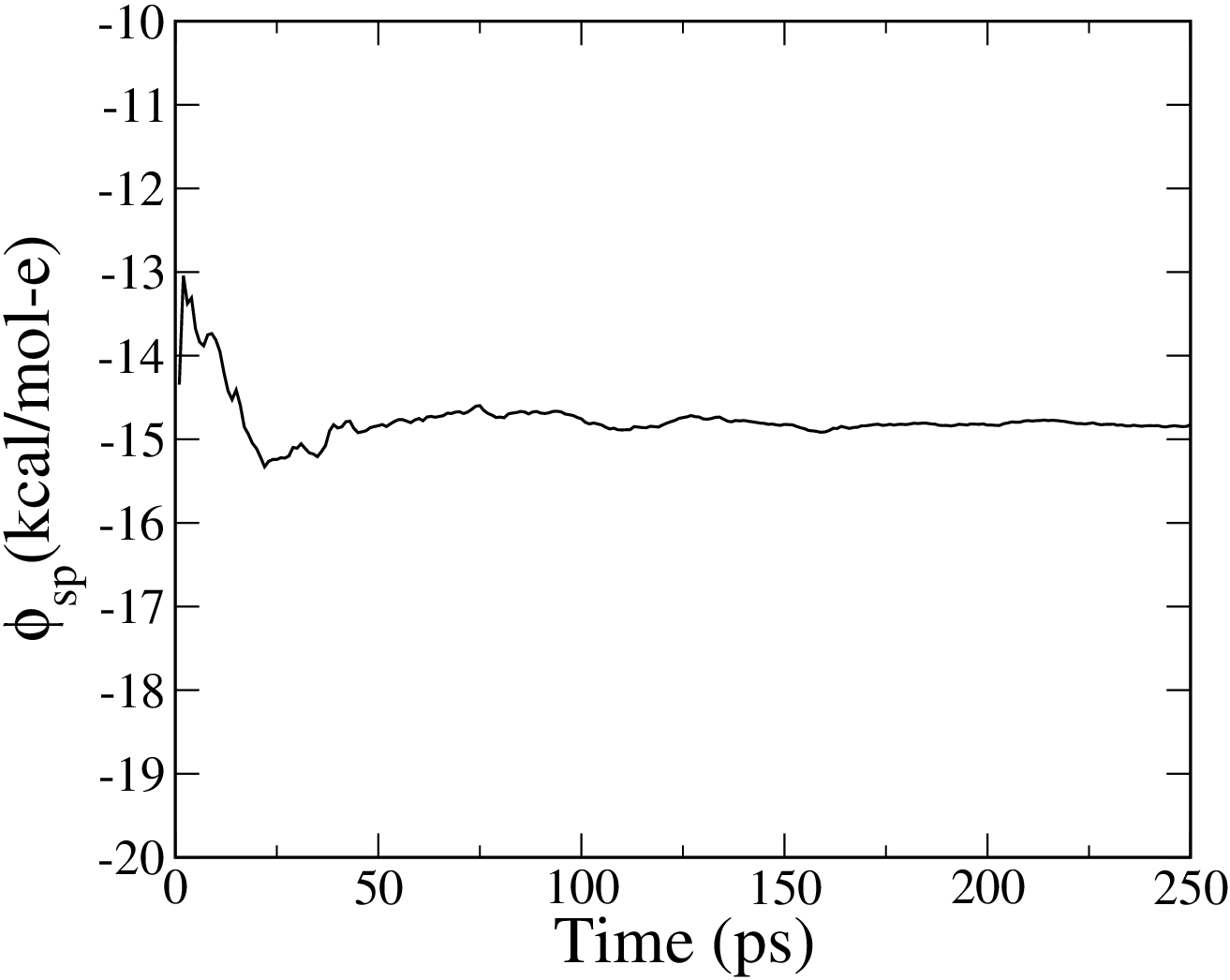} 
\caption{The average surface potential as a function of time (ps) computed for the SPC/E $n=511$ cluster using the 
far-field Ewald potential.}
\par\end{center}
\label{fig:spvst}
\end{figure}

A similar strategy can be imagined for analysis of the surface potential
obtained from an AIMD simulation. In periodic boundaries, the far-field term is handled in ${\bf k}$-space, as was done in the classical simulations of Ref.~\cite{tlbsurf}. The Ewald parameter $\eta$ can be
chosen so that the majority of interactions with distant nuclei
occurs outside the range of the nonlocal part of the pseudo-potential,
and thus the effective charge for each nucleus is simply the total
charge minus the charge of the omitted core electrons. Thus any potential ambiguity 
involving the pseudo-potential is also removed.

\section{Results}

Results are first presented concerning classical point-charge models of the I$^-$ ion in $n=105$ and $n=511$ SPC/E water droplets. The LMFT length scale parameter was chosen as $\eta^{-1} = 5.0$~{\AA}. The radial water density profiles for the uncharged and fully coupled ions are shown in Fig.~2. The solute particle was restrained to lie at the cluster center of mass.  It is clear from these plots that turning on the charge leads to a substantial rearrangement of the local hydration structure; the density profile near the liquid-vapor interface is not appreciably altered by the transition, however.  

\vspace{.2cm}

\begin{figure}
\begin{center}
\includegraphics[scale=0.3]{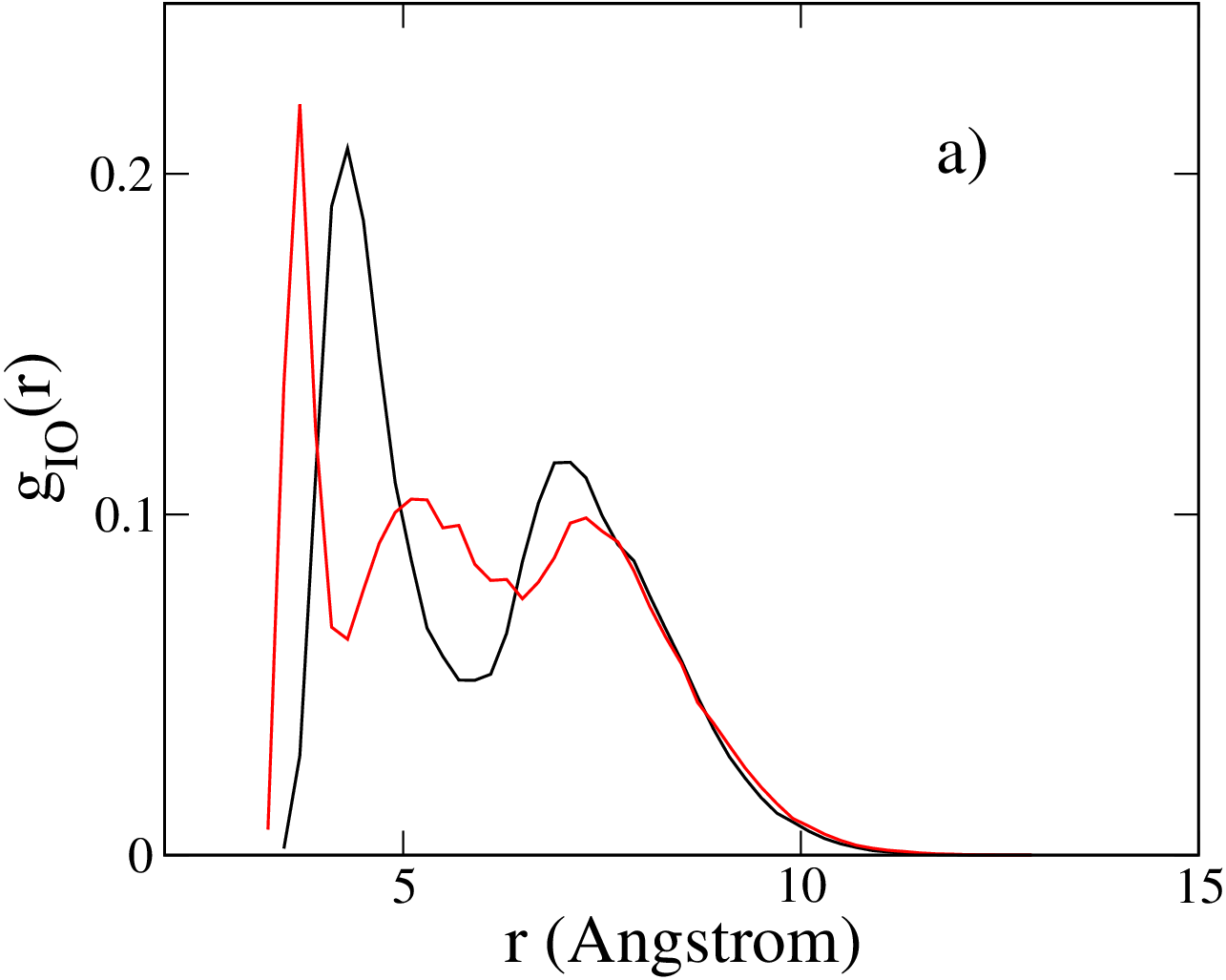} 
\includegraphics[scale=0.3]{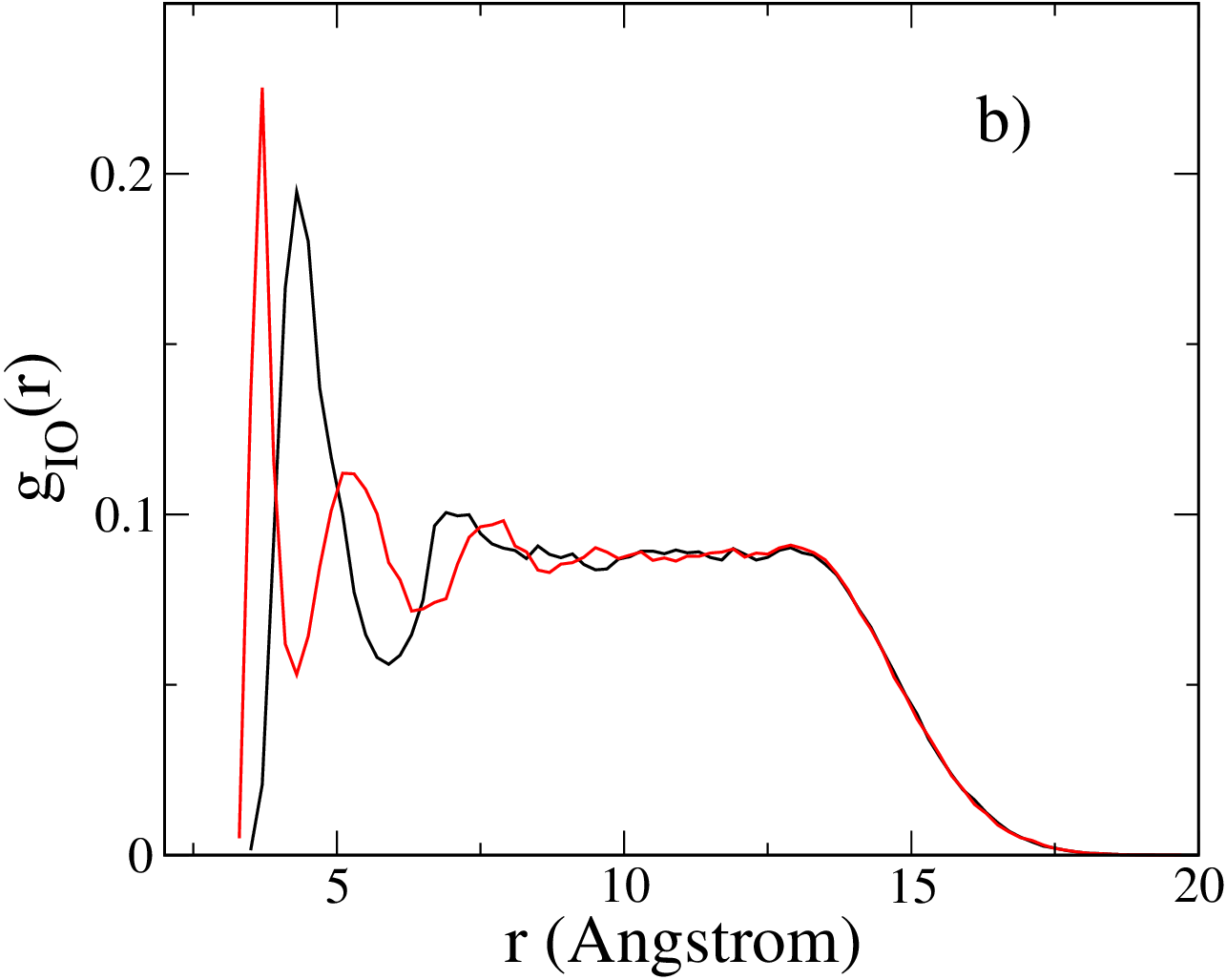}
\caption{The water oxygen radial density profiles (arbitrary units) for the $n=105$ (a) and $n=511$ (b) clusters, computed with the ion uncharged (black) and charged (red).}
\par\end{center}
\label{fig:rhor}
\end{figure}

Initially, the $n=105$ cluster was simulated at 250 K in order to prevent evaporation events from the surface.  The freezing 
point for a cluster is substantially lower than the bulk value, and the freezing temperature of the SPC/E model is 214 K \cite{spce-freeze}.  Visual inspection of animations of the cluster motion show clearly that it is in a liquid state at 250 K.  Subsequent simulations were performed at 300 K, resulting in a nearly identical water density profile with little or no evaporation.  As discussed below, the surface potential and free energy properties of the cluster did not change appreciably between 250 K and 300 K.  The quantum calculations reported below were performed primarily on configurations taken from the 250 K $n=105$ simulation, with one set of calculations on the $n=242$ cluster at 300 K.

\begin{table}
\caption{Data for the surface potential and free energy calculations.  All energies are in kcal/mol and all potentials
are in kcal/mol-e.   All errors are on the order of 0.1-0.2 kcal/mol or kcal/mol-e, except the QM surface potential calculation, with an estimated error of 0.7 kcal/mol-e. The value for the vdW free energy contribution was taken from Ref.~\cite{tlbsurf} (slab geometry).  The vdW value at 250 K was shifted downward by 0.6 kcal/mol  since the hydration entropy of the vdW particle is roughly -12.7 cal/mol-K. The finite-size dielectric correction for the distant interactions with water assumed cluster radii of 9.0~{\AA} for the $n=105$ cluster and 15.0~{\AA} for the $n=511$ cluster, resulting in corrections of $-18.2$ kcal/mol and $-10.9$ kcal/mol, respectively. All results are for the particle constrained to the cluster center of mass, except for those labeled by `surf', which are results for the neutral particle located 8.5~{\AA} from the cluster center for the $n=105$ cluster. The bulk periodic boundary free energy (from Ref.~\cite{tlbsurf}) is shifted by $-q\phi_{sp}$ as implied in Eq.~\ref{eq:gibbs}.}
\begin{center}
\begin{tabular}{lcccc}
\hline
& \multicolumn{1}{c}{CM potential data} \\
\hline
Geometry & $\phi_{sp}$ & $\phi_{lp}$ & $\phi_{np}$& \\
\hline
$n=105$ (250 K, CM) & -14.1 & +6.6 & -7.5 & \\
$n=105$ (250 K, CM, surf) &  &  & -3.6 & \\
$n=105$ (300 K, CM) & -14.0 & +6.9 & -7.2 &\\
$n=511$ (300 K, CM) & -14.0 & +7.5 & -6.5 &\\
Bulk Slab (300 K, CM) & -13.8 & +8.1 & -5.7 &\\
No solute $n=105$ (300 K, CM) & -14.7  & & &\\
No solute $n=511$ (300 K, CM) & -14.9  & & &\\
\hline
& \multicolumn{1}{c}{QM potential data} \\
\hline
$n=105$ (250 K, QM) & +67.0 & -76.9 & -9.9 & \\
$n=105$ (250 K, QM, surf) &  &  & -3.8 & \\
\hline
& \multicolumn{1}{c}{Free energy data} \\
\hline
Geometry & $\mu_{vdW}^{ex}$ & $\mu_{es,loc}^{ex}$ & $\mu_{es,far}^{ex}$& $\mu^{ex}$ \\
\hline
$n=105$ (250 K, CM) & 3.9 & -30.4 & -14.6 & -59.3 \\
$n=105$ (300 K, CM) & 4.5 & -31.5 & -14.5 & -59.7 \\
$n=511$ (300 K, CM) & 4.5 & -31.9 & -20.3 & -58.6\\
Bulk slab (300 K, CM) &  &  &  & -58.3\\
Bulk PBC (300 K, CM) &  &  &  & -71.5 \\
\hline
\end{tabular}
\end{center}
\end{table}

The classical results for the potential at the center of the uncharged ion $\phi_{np}$ are first presented (Table 1).  The results show that the classical surface potentials calculated in the cluster and slab geometries are consistent (with a shift to more negative values by 1.1 kcal/mol-e or 0.05 V), even for a cluster as small as $n=105$, a case where a flat water radial density profile is not obtained near the surface.  The cluster results with the uncharged solute included are shifted slightly to less negative values relative to the data presented above for the case of no solute present.  This small shift is likely caused by local rearrangements of waters in the first hydration shell; the far-field potential smoothly grows in magnitude with increasing radius and thus a small contribution from that local region can lead to a small shift.  The local potential $\phi_{lp}$ is positive and increases somewhat with increasing cluster size. Nevertheless, it is clear that the cluster calculations capture the essential physics of the potential at the center of the neutral particle, with a slightly more negative net potential compared with the previous bulk slab calculations.   

It can also be noted that, from the $n=105$ simulations at the two temperatures, there is only a very weak temperature dependence to both $\phi_{sp}$ and $\phi_{lp}$ (and the resulting $\phi_{np}$), with most of the change coming from $\phi_{lp}$.  The results above show clearly that a sizable surface or net potential can exist, even with very small temperature derivatives of the potentials. The same conclusion was previously reached by Sokhan and Tildesley \cite{tildesley_wsp}.

Utilizing the LMFT partitioning methods developed in Refs.~\cite{tlbloc11,tlbent11,tlbsurf}, the computed hydration free energies are also shown in Table 1. The computed excess chemical potentials at the cluster center include one finite-size correction, which is a Born model correction that extrapolates to the large-cluster limit ($\mu_{diel}^{ex} = - (1 - 1/\epsilon) q^2/2R_D$, where $R_D$ is the droplet size).  That dielectric correction includes no contribution from the surface potential (which is already in the cluster result),  and is expected to be accurate for handling the distant interactions with water.  The resulting free energies are in excellent agreement with the experimental value of $-59.3$ kcal/mol \cite{mtiss98}.  It is encouraging that insertion of the ion into a cluster as small as $n=105$ can result in a quite accurate free energy estimate.  Importantly, these results clearly include contributions from the surface potential, and the agreement with experiment thus suggests that the tabulation of Ref.~\cite{mtiss98} does include that effect for all of the ion free energy values. We also note that, in agreement with previous results \cite{tlbsurf}, the far-field electrostatic contribution is accurately Gaussian in the cluster geometry.  

It is also interesting that, for the $n=105$ cluster, the local electrostatic contribution to the free energy becomes {\it less negative} at lower temperature.  This free energy change implies a {\it positive} local electrostatic contribution to the entropy, an effect previously observed in Ref.~\cite{tlbent11}.   There it was suggested that the positive value arises from a competition between hydrophobic and hydrophilic hydration for the large (chaotropic) anion.  Consistent with Ref.~\cite{tlbent11}, the results also provide further confirmation that the fluctuation contribution to the entropy is relatively local.  From the results for the total free energy change between the two temperatures, a rough estimate of $+1.5$ cal/mol-K results for the hydration entropy (including the negative non-polar contribution).  This very small, positive result compares well with the value obtained experimentally (0.2 cal/mol-K) in Ref.~\cite{mtiss98}.  Using different ion models, Ref.~\cite{tlbent11} obtained a result of $-0.3$ cal/mol-K for the I$^-$ ion, and the results were compared to the tabulation in Ref.~\cite{rschm00}, exhibiting some deviation. Since the measured results in Ref.~\cite{mtiss98} are likely of higher accuracy, we can consider the agreement encouraging. These results illustrate the strong ion specificity in hydration entropies, since small, kosmotropic ions display negative entropies of large magnitude \cite{tlbent11}.  

For comparison, the hydration free energy of the Na$^+$ ion in the $n=511$ cluster was also computed, with a result of $-105.8$ kcal/mol. This result is shifted from the experimental value of $-103.2$ kcal/mol \cite{mtiss98}. The resulting computed NaI total is $-164.4$ kcal/mol, compared with the unambiguous experimental total of $-162.5$ kcal/mol \cite{mtiss98}.  It can be noted that the force fields developed in Refs.~\cite{horinek_rational_2009} and \cite{horinek_specific_2009} used a water surface potential value of $-12.1$ kcal/mol-e, which is somewhat smaller than the computed value of $-13.8$ kcal/mol-e for the SPC/E model \cite{tlbsurf}. The force fields were fit to ion hydration free energies and entropies in reference to the experimental results of Ref.~\cite{mtiss98}. Thus, the parametrized force fields may reflect that difference.  Also, calculations in the cluster geometry may differ mildly from bulk or slab geometries (see discussion below).  If both I$^-$ and Na$^+$ results are shifted upward by 0.9 kcal/mol to reflect the pair discrepancy, final values of $-57.7$ and $-104.9$ kcal/mol are obtained, respectively.  Compared with experiment these values are shifted up by 1.6 kcal/mol and down by 1.7 kcal/mol, respectively. These symmetric shifts likely are a direct reflection of the slightly larger surface potential magnitude observed for the clusters {\it vs.}~the slab geometry. 

In order to obtain a rough estimate of the influence of the various terms in Eq.~\ref{eq:pdtnp} on the driving forces for ions near the water surface, each electrostatic term was computed for the $n=105$ cluster at 300 K for the I$^-$ situated both at the cluster center and at a distance of 8.5~{\AA} from the center (near the dividing surface).  The nonpolar contribution was estimated from the slab data in Ref.~\cite{tlbsurf}. We consider changes in four terms: the nonpolar contribution, the net potential term, the binding energy difference term, and the fluctuation term. The resulting free energy changes are -4.2, -4.2, +10.2, and +0.9 kcal/mol, respectively, leading to a net repulsion of +2.7 kcal/mol. (Ref.~\cite{horinek_specific_2009}, Fig.~10 iodide set 2, shows that the free energy profile, using these classical models, begins its rise to positive values inside the dividing surface, and the radius chosen here was at the dividing surface for the cluster, leading to the positive free energy change.)   The nonpolar and net potential portions both contribute a driving force towards the surface, while the binding energy term yields a large repulsion. This repulsion is due both to the loss of part of the hydration shell locally, and reduction in the distant interactions as well.  The fluctuation term yields only a small repulsion.  These calculations were redone in the slab geometry, with the following results (listed in the same order): -4.2, -2.6, +7.2, and +0.5 kcal/mol (the near-surface position for these calculations was chosen 0.5 {\AA} inside the dividing surface). Again, the change in the average fluctuation term is small.  

Finally, quantum mechanical (DFT) results are presented for the distribution of the net potential, $\phi_{np}$, and the results are compared with the classical distribution.  Configurations (1000 in total) were drawn from a 5 ns classical simulation of the $n=105$ cluster at 250 K, as was done in Ref.~\cite{leung_surface_2010}. The quantum net potential at the center of the neutral particle was computed for each configuration using the ORCA code (DFT-BLYP, cc-pVDZ basis).  The distributions were also computed with the particle restricted to a radius of 8.5~{\AA} from the center of mass to examine the electrostatic environment near the cluster surface.  Fig.~3 displays the results.  

It should be noted here that the classical particle was removed from the water cluster for computation of the net potential. Thus the computed potential is that for an ion-sized cavity, and is not exactly the same as $\langle \varepsilon \rangle_0$ in Eq.~\ref{eq:pdtnp}, which is the interaction energy of the ion with the surrounding waters sampled with no charge on the ion.  The simple rearrangement into the form $\langle \varepsilon \rangle_{cav} + (\langle \varepsilon \rangle_0-\langle \varepsilon \rangle_{cav})$ then shows that the correct evaluation of $\langle \varepsilon \rangle_0$ involves the cavity term plus the change in the local part of the potential (the far-field parts canceling) incurred by turning the point charge into the ion.  Thus $\langle \varepsilon \rangle_{cav}$ calculated here is a first estimate of the net potential $\phi_{np}$. The physical assumptions are that the ion charge distribution is spherical, there is no overlap with the water electron density, and only electrostatic interactions are included.  In reality, the ion charge density will be distorted due to the nearby waters and there will be some electron density overlap. This ambiguity does not arise in the classical calculation. 

\begin{figure}
\begin{center}
\includegraphics[scale=0.4]{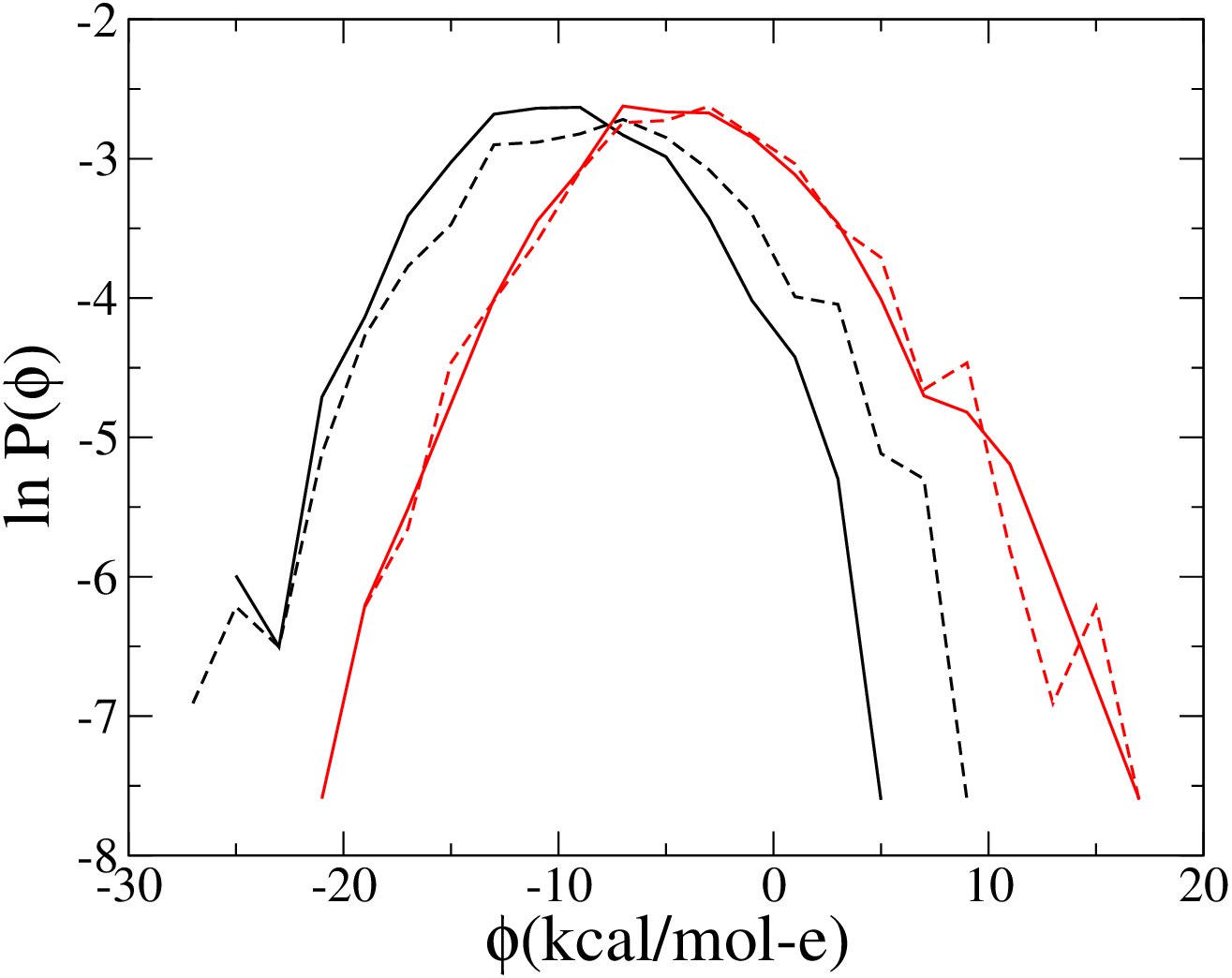}
\caption{Net potential distributions at the center computed with classical and quantum models for the $n=105$ cluster.  The black curves are for the cavity located at the cluster center, while the red curves are for the cavity near the cluster surface. The full curves are the quantum (DFT-BLYP, cc-pVDZ basis) results and the dashed curves are the classical SPC/E results. } 
\par\end{center}
\end{figure}

For the particle situated both at the cluster center and near the surface, the distributions of the cavity potential are relatively consistent between the quantum and classical SPC/E results. 
(A recent study suggests a classical model dependence; the TIP5P model results in a near-zero average net potential \cite{mundy-foot1}.)  As shown in Table 1, the average net potential for the particle at the cluster center is shifted downward by 2.4 kcal/mol-e for the quantum calculation compared with the classical SPC/E result (to a value of -9.9 $\pm 0.2$ kcal/mol-e). This result suggests that the negative net potential is indeed maintained when modeling water at the quantum mechanical level.  The classical and quantum net potentials computed at the cluster surface are very close to each other, with the average quantum result for the net potential shifted downward by only 0.2 kcal/mol-e.  It can be noted that, while the quantum and classical distributions are very similar near the cluster surface (and skewed in the positive direction), the classical distribution exhibits a shift and a broader spread than the quantum distribution at the droplet center. Any asymmetry in the distributions is due to local interactions, since the far-field contribution is Gaussian distributed \cite{tlbsurf}. Also, the range of the distributions is large, spanning 35 kcal/mol-e or 1.5 V; thus aggressive sampling (5 ns classical simulation here) may be an important factor.  Collecting the results,  the net potential changes moving from the cluster center to the surface are $+3.9$ kcal/mol-e (classical) and $+6.1$ (quantum).  The results imply a net driving force pushing anions toward the surface and cations away.  

Followup calculations were performed to test the robustness of the computed cavity potential.  Using the same basis and cluster size, DFT-B3LYP and Hartree-Fock calculations resulted in values of $-10.0 \pm 0.5$ kcal/mol-e and $-8.7 \pm 0.5$ kcal/mol-e, respectively. A DFT-BLYP calculation with the Aug-cc-pVDZ basis set (that includes diffuse functions) yielded a value of $-7.4 \pm 0.5$ kcal/mol-e.  Finally a DFT-BLYP calculation (using the cc-pVDZ basis set) for an $n=242$ water cluster resulted in a cavity potential of $-7.9 \pm 0.7$ kcal/mol-e.  Finally, one calculation was performed at the DFT-BLYP level in which the classical particle was sampled with a purely repulsive (WCA) potential as opposed to the full Lennard-Jones potential. The computed cavity potential for that case is $-9.2 \pm 0.5$ kcal/mol-e.  

Of course the computed quantum net potentials can depend on the sampling employed here, namely using the classical SPC/E model over long times.  This choice of sampling was chosen in order to allow for robust sampling of a diverse set of independent configurations for subsequent analysis.  There is ample evidence, however, from both experiment \cite{cremer-spcevsexpt} and {\it ab initio} simulations \cite{mundy-aimdvspc,kaxiras-aimdvspc} that the SPC/E model accurately reproduces the orientational distributions of the water molecules near the liquid-vapor interface.  Another observation is that, assuming the net potential results from the relative water dipole orientations at the bulk and solute-solvent interfaces, the SPC/E result of -7.5 kcal/mol-e (for the $n=105$ cluster) produces the quantum mechanically computed result of -9.9 kcal/mol-e when the classical result is scaled by the ratio 3.1/2.35; this is the ratio of the water dipole moments estimated from {\it ab initio} simulations \cite{eguar09} and the SPC/E dipole moment (2.35 D).  Presumably, one component of the difference between the quantum result computed here (-9.9 kcal/mol-e) and that estimated from the LPS analysis above (-11.6 kcal/mol-e) is the above-noted difference between the cavity potential and the $\langle \varepsilon \rangle_0$ quantity discussed above. 

In a recent {\it ab initio} CP2K-DFT simulation study (BLYP-D and PBE-D functionals) examining the cavity potential in the bulk slab geometry \cite{mundy-foot1}, average potentials very close to 0 or slightly positive were obtained, differing from the results presented here. Differences between these two sets of calculations include: cluster vs.~slab geometries, all-electron vs.~pseudopotential calculations, Gaussian vs.~dual Gaussian/plane-wave basis sets, and the means of sampling (classical SPC/E over 5 ns time scale vs.~AIMD over 20-30 ps).  Presently, the origin of the difference in the net potentials is not known.  One source of potential concern is the incorrect asymptotic form for the potential in the DFT-BLYP model (and other gradient-corrected functionals) \cite{baerends}; since the net potential results from the molecular dipole contributions of two different interfaces, and the potential may depend sensitively on the tails of the electron distributions, this issue deserves further investigation.  As a preliminary step, the Hartree-Fock results presented here possess the correct asymptotic behavior.  A result of approximately zero for the net potential would be somewhat surprising, since this implies an exact cancellation of the molecular dipole contributions between the bulk surface and the boundary between the small solute and water.  Leung \cite{leung_surface_2010} observed a close correspondence between the SPC/E and DFT results for the dipole contribution to the surface potential, but both models yielded a very small magnitude value due to the small system size (slab geometry).  Naively, one might expect a higher dipole density near the uncharged solute due to the increased water density there. The classical (TIP4P-FQ) molecular dipole surface potential contribution profile in Ref.~\cite{warren_hydration_2007} is positive, and if the dipole density is somewhat higher near the solute, then a net negative potential at the cavity center would be predicted, as we see in the SPC/E model. 


Finally, results are presented for the quantum local and far-field contributions to the potential at the cavity center.  To compute the far-field contribution, the charge density was obtained for 21 statistically independent configurations and written to cube files.  The grid spacing for the electron density was taken as 0.08 au.  On this grid, integration of the electron density led to slight deviations (on the order of 0.05 e) from a total charge of $-1050$.  Thus all of the nuclear charges were scaled to yield a total positive nuclear charge that exactly balances the calculated electronic charge.   Then, by comparison with individual net potentials computed with the ORCA code, the net potentials agreed to better than 1 kcal/mol-e. It is expected that the deviations in the net potentials arise mainly from local contributions, so these results suggest the far-field values computed on the grid are of high accuracy.  The resulting average far-field potential is $+67.0 \pm 0.7$ kcal/mol-e ($+2.9$ V).  Leung \cite{leung_surface_2010} has shown that the vast majority of this large value comes from the molecular quadrupole contribution. 

Two previous DFT studies of the water surface potential in the slab geometry have obtained values of $+3.1$ V \cite{kathmann_understanding_2011} and $+3.6$ V \cite{leung_surface_2010}.  The value reported here for the cluster geometry shows modest agreement with these estimates, but is slightly smaller in magnitude.  The slight discrepancy likely arises due to the use of a modest-sized basis set in the present calculations.  As discussed above, the potential jump across the interface involves a contribution from the trace of the molecular quadrupole moment.  The quadrupole trace for the SPC/E model is $+8.4$ au (taking the center of mass as the origin) and the dipole moment is 2.35 D. The quantum mechanical trace for a single water molecule in the SPC/E geometry, using the BLYP functional and the cc-pVDZ basis is -38.1 au, while the dipole moment is 1.79 D (compared with the experimental value of 1.85 D).  
Using a higher-level aug-cc-pVTZ basis leads to a quadrupole trace of $-42.3$ au for the same geometry and a dipole moment of 1.74 D. From these results it is clear that the large difference in $\phi_{sp}$ computed classically and quantum mechanically arises mainly due to the different signs and magnitudes of the water quadrupole moments in the point-charge and quantum models.  A recent study has discussed the importance of the large quadrupole of the water molecule \cite{ichiye}, and alternative classical multipole and smeared-charge based water models can accurately reproduce the quantum quadrupole moment \cite{burke1,burke2,cummings-w}.

If the SPC/E structure is then minimized, and the aug-cc-pVTZ basis is used, the quantum quadrupole trace is $-42.3$ au and the dipole moment is 1.81 D.  Calculations at the MP2 level for the relaxed geometry, using the same high-level basis set, result in virtually the same quadrupole trace and a dipole moment of 1.86 D.  Thus we can see that the basis set used here underestimates the quadrupole trace by roughly 11\%.  Relaxing to the minimum energy geometry does not substantially alter the quadrupole trace but does modify the dipole slightly.  Of course in the condensed phase, water molecular polarization can lead to changes in both the dipole and quadrupole.   If we simply scale the observed result of 2.9 V by 1.1 (since the quadrupole contribution completely dominates the dipole contribution), however,  the final result of 3.2 V is obtained, in decent agreement with the previous bulk results.  The present cluster results are computed at the all-electron level and do not involve Ewald calculations of the potential.  The agreement with the bulk results is therefore encouraging and the results provide further confirmation of the large positive value of the surface potential when computed quantum mechanically.  The present result also can be compared with the value measured in electron holography experiments ($+3.5$ V) \cite{holography}. 

Taken together, the quantum $\phi_{np}$ result of $-9.9$ kcal/mol-e and the $\phi_{sp}$ result of $+67.0$ kcal/mol-e imply a 
$\phi_{lp}$ value of $-76.9$ kcal/mol-e.   These results illustrate the large magnitude of the quadrupole contribution, as was previously shown for the bulk slab geometry \cite{leung_surface_2010}, and provide further circumstantial evidence for the complete cancellation of this contribution.   

\section{Summary Discussion}

A principal aim of the present paper is to attempt to reconcile the now well established surface potential at the water liquid-vapor interface (3.1 to 3.6 V) with electrochemical estimates that fall in the range $-0.4$ to $+0.3$ V.   The resolution comes through the rearrangement of Eq.~\ref{eq:gibbs} for the real hydration free energy into Eq.~\ref{eq:gibbs1}.  The same analysis has been previously applied to classical studies to exhibit the cancellation of model-dependent quadrupole contributions that arise in the local and surface potential terms that sum to the net potential \cite{harder_origin_2008,allen_SP}.  The present paper has extended that analysis to quantum mechanical calculations. The quantum calculations are required for comparison to experimental surface potential measurements due to the very different classical and quantum water quadrupole moments.  But we can conclude, based on the cancellation involved in Eq.~\ref{eq:gibbs1}, that the water molecular quadrupole contribution to the interface potentials does not play a role in ion hydration thermodynamics.  

While Eq.~\ref{eq:gibbs} is a correct expression of the electrochemical potential, it was shown above that the resulting intrinsic hydration free energies differ widely from any sensible bulk hydration free energy estimates reported in the various tabulations.  It is suggested here that the re-analysis of the Latimer, Pitzer, and Slansky (LPS) model \cite{latimer_39} by Ashbaugh and Asthagiri \cite{ashbaugh_single_2008} provides the necessary connection to relate the disparate surface potential values arising from electron holography and electrochemical measurements.  The single-ion LPS results obtained by fitting Eq.~\ref{eq:lps} are a close approximation to the bulk hydration free energy defined above, $\mu_b^{ex}$. 

If we compare the resulting bulk free energies to those measured in Ref.~\cite{mtiss98}, with a charge-dependent shift of magnitude 11.6 kcal/mol-e,  excellent agreement is obtained for both the cation and anion series (to within less than 1 kcal/mol for each ion). We take this as an experimentally based estimation of $\phi_{np} = -11.6$ kcal/mol-e ($-0.50$ V). Due to the consistency across the cation and anion series, we can thus consider the derived $\phi_{np}$ value to be mainly a property of water itself (at the bulk surface and near the spherical solute, with a small solute size dependence). Considering the relatively low level of the quantum mechanical calculations employed here, the cavity potential result obtained above, $-9.9$ kcal/mol-e (or $-0.43$ V), agrees modestly well with the LPS result.  The results further suggest a $\mu_b^{ex}$ value for the proton of $-254.3$ kcal/mol to go with the real hydration free energy of $-265.9$ kcal/mol. This $\mu_b^{ex}$ value for the proton matches {\it exactly} the value  given by Marcus \cite{marcus} (when corrected for the standard state). (His value was determined using a method developed by Halliwell and Nyburg \cite{halliwell}; that method fits deviation of cation and anion free energies to a power series in the ion radius, starting with the third power, and extrapolating to large ions.)  

A completely different approach to single-ion free energies was taken by Schmid, Miah, and Sapunov \cite{rschm00}.  Their analysis rests on the assumption of equal aqueous entropies for the proton and hydroxide ions; the data in Ref.~\cite{mtiss98} result in a difference of less than 3 cal/mol-K for these two ions, supporting that assumption.  Following that assumption, their method uses only bulk pair hydration free energy data, and thus includes no interfacial potential contributions.  Their result for the proton hydration free energy is $-251.4$ kcal/mol, shifted by 14.5 kcal/mol from Ref.~\cite{mtiss98} (indicating $\phi_{np} = -14.5$ kcal/mol-e). Alternatively the shift in the proton hydration enthalpy estimated from the TATB hypothesis yields a predicted net potential of $-11.0$ kcal/mol \cite{marcus_tatb}. As can be seen, this relatively wide range of approaches yields a consistent indication of a shift between real and bulk hydration free energies, indicating a negative net potential $\phi_{np}$. Further work is needed, however, to go beyond the estimation of the cavity potential presented here and accurately determine the $\left<\varepsilon_{es} \right>_0$ quantity in Eq.~\ref{eq:pdtnp}. 
Several recent papers have addressed the issue of the absolute free energy of hydration of the proton \cite{dixon_proton,asthagiri_absolute_2003,agros03,glamo06,kelly_aqueous_2006,goddard,fawcett08}. It is hoped that the present paper will help clarify the various discussions in those works concerning the role of the surface potential. 

The $\phi_{np}$ result discussed above differs from the conclusion of Randles \cite{randles77} based on the temperature dependence of cell emfs (the measured temperature derivative is small in magnitude and negative). Interpretation of those experiments assumes a smooth extrapolation to the critical point, leading to very small positive estimates of the surface potential.  As noted in the results above, a small temperature derivative to the surface potential does not necessarily imply a small value for the surface potential itself, however. This point was previously noted by Sokhan and Tildesley \cite{tildesley_wsp}. Thus we consider the negative value for $\phi_{np}$ obtained here as more convincing since it allows for reconciliation of bulk thermodynamic and recently obtained cluster data.  

In terms of implications for ion specificity in hydration, the resulting $\mu_b^{ex}$ values for the K$^+$ and F$^-$, which are of comparable size, are $-74.4$ and $-116.0$ kcal/mol, respectively.  These values indicate a distinct charge asymmetry and show that anions are more strongly hydrated for a given ion size. Another possible implication relates to the recent debate concerning the relative propensities for the proton and the hydroxide ions for the water surface \cite{jungwirth-acid,beattie,saykally-acid}.  The measured isoelectric point for the surface is in the pH range of 2-4 \cite{beattie}; if an average potential of roughly half the $\phi_{np}$ values above is assumed in the layer near the surface, the predicted protonation shift for water would fall in this range. 

Another conclusion of the present paper is that the influence of the water surface potential on ion distributions is related to the above discussion of bulk hydration free energies, and is not a separate issue \cite{tlbsurf,mundy-SP-12}. In fact, a part of the driving force selecting for anions at the water surface arises from exactly the same $q\phi_{np}$ free energy term.  Eq.~\ref{eq:pdtnp} makes the necessary connection, showing that the negative net potential provides a net driving force that moves anions to the surface while repelling cations. In colloquial terms, the ions never experience the full surface potential $\phi_{sp}$, but rather $\phi_{np}$, near the interface. Previous classical simulation results showing ions moving back into the bulk when the charge is changed from negative to positive support this view \cite{swedes,vaitheeswaran_hydrophobic_2006,sagui05}.  In addition, if the net potential effect were ignored, the Na$^+$ ion would be preferred at the water surface over the I$^-$ ion \cite{tlbsurf}. The net potential $\phi_{np}$ is due to long-ranged interactions, arising from the difference of surface and solute-solvent boundary dipole layers, but generates a driving force on ions only within less than 10 {\AA} from the dividing surface. Of course other more local factors are crucial too in determining the details of the ion distributions near the surface. Taken together, the analysis of the hydration free energy shifts and surface-potential-related driving forces provide a more unified picture of ions near the interface.   These conclusions leave one outlying case, however, which is that the F$^-$ ion is repelled from the surface \cite{jungwirth}, even with the driving force arising from the surface potential. The repulsion arises from detailed local interactions that differ near the surface, and will be the subject of a future paper.  

To close, a few observations are made concerning the results in the present study. One is that the
very simple classical point charge models (and more accurate polarizable models) can yield sensible physical results for ions near the water surface, although the results are of course not identical to the quantum models.  The results suggest that the potential outside of the classical water models can be relatively accurate in an average sense, while the charge distribution inside the molecules is completely unphysical \cite{kathmann_understanding_2011}.  Recent results suggest, however, a model dependence to the cavity potential between the SPC/E and TIP5P classical models (see discussion above).  

The free energy results obtained from the classical droplet models are surprisingly robust, and provide further indications that the classical models represent a large fraction of the important physics of ion hydration. Neglected are more accurate representations of dispersion interactions 
\cite{wkunz04,barrybook}, polarization \cite{jungwirth,chang_recent_2006}, and chemical effects such as charge transfer \cite{zhao10,collins-rev-12}, effects that may have important impacts.  A suggestion, perhaps heretical, is that modeling free energies in the cluster geometry may provide a more physically grounded approach for ion force-field development.  One reason is that, instead of multiple periodic-boundary finite-size corrections \cite{hunenberger-cont,hunenberger_3,ourbook}, there is only one free energy correction for the interaction of the ion with distant waters.  This correction is likely accurate so long as the droplet size is greater than roughly 10~{\AA}. Another possibility is a general intuition that the physical environment at the center of a droplet may be more `natural' than that in periodic boundaries.  Observed shifts of {\it local} electrostatic contributions passing from the bulk (no interfaces) to the slab geometry provide a hint of slight structural rearrangements near the ion with the introduction of free interfaces \cite{tlbsurf}.   

The present paper has focused on single ions in water droplets.  The discussion has suggested that the water surface potential is included in the measurements of Ref.~\cite{mtiss98}, that the net potential rationalizes observed shifts compared with bulk estimates, and that the net potential influences ion distributions near the surface.  A previous paper by Zhou, Stell, and Friedman \cite{zsf}, carried further by Pratt \cite{pratt_contact_1992}, provides a very different picture for the contact potential between two (conducting) bulk phases, each of which contains some (perhaps small) density of ions.   Those papers show that the contact potential $\phi_{sp}$ is actually determined by the differences of the {\it intrinsic} hydration free energies of the ions between the two phases (due to charge neutrality in the two phases). The derivation can also be performed starting from Eq.~\ref{eq:gibbs1}, leading to an expression for the net potential $\phi_{np}$ in terms of the bulk hydration free energies $\mu_b^{ex}$ discussed above. The net potential is not strictly ion-independent, but shows only small variation with ion size.  Thus, as opposed to the discussion above focused on the effect of the surface potential on the ion distributions, the bulk ion hydration thermodynamics determines the surface or net potential.   Further discussion of these important results is outside the scope of this paper, but the conclusions illustrate the extreme subtlety involved in ion hydration thermodynamics.  

\section{Acknowledgments}

I would like to thank Lawrence Pratt, Chris Mundy, Marcel Baer, Greg Schenter, Shawn Kathmann, Dilip Asthagiri, David Rogers, Dominik Horinek, and Keven Leung for helpful comments.  This research was supported by NSF grant CHE-1011746 and a generous grant of computer time at the Ohio Supercomputer Center.

\bibliographystyle{apsrev4-1}
\bibliography{tlbsp}

\begin{thebibliography}{97}%
\makeatletter
\providecommand \@ifxundefined [1]{%
 \@ifx{#1\undefined}
}%
\providecommand \@ifnum [1]{%
 \ifnum #1\expandafter \@firstoftwo
 \else \expandafter \@secondoftwo
 \fi
}%
\providecommand \@ifx [1]{%
 \ifx #1\expandafter \@firstoftwo
 \else \expandafter \@secondoftwo
 \fi
}%
\providecommand \natexlab [1]{#1}%
\providecommand \enquote  [1]{``#1''}%
\providecommand \bibnamefont  [1]{#1}%
\providecommand \bibfnamefont [1]{#1}%
\providecommand \citenamefont [1]{#1}%
\providecommand \href@noop [0]{\@secondoftwo}%
\providecommand \href [0]{\begingroup \@sanitize@url \@href}%
\providecommand \@href[1]{\@@startlink{#1}\@@href}%
\providecommand \@@href[1]{\endgroup#1\@@endlink}%
\providecommand \@sanitize@url [0]{\catcode `\\12\catcode `\$12\catcode
  `\&12\catcode `\#12\catcode `\^12\catcode `\_12\catcode `\%12\relax}%
\providecommand \@@startlink[1]{}%
\providecommand \@@endlink[0]{}%
\providecommand \url  [0]{\begingroup\@sanitize@url \@url }%
\providecommand \@url [1]{\endgroup\@href {#1}{\urlprefix }}%
\providecommand \urlprefix  [0]{URL }%
\providecommand \Eprint [0]{\href }%
\providecommand \doibase [0]{http://dx.doi.org/}%
\providecommand \selectlanguage [0]{\@gobble}%
\providecommand \bibinfo  [0]{\@secondoftwo}%
\providecommand \bibfield  [0]{\@secondoftwo}%
\providecommand \translation [1]{[#1]}%
\providecommand \BibitemOpen [0]{}%
\providecommand \bibitemStop [0]{}%
\providecommand \bibitemNoStop [0]{.\EOS\space}%
\providecommand \EOS [0]{\spacefactor3000\relax}%
\providecommand \BibitemShut  [1]{\csname bibitem#1\endcsname}%
\let\auto@bib@innerbib\@empty
\bibitem [{\citenamefont {Harscher}\ and\ \citenamefont
  {Lichte}(1998)}]{holography}%
  \BibitemOpen
  \bibfield  {author} {\bibinfo {author} {\bibfnamefont {A.}~\bibnamefont
  {Harscher}}\ and\ \bibinfo {author} {\bibfnamefont {H.}~\bibnamefont
  {Lichte}},\ }\href@noop {} {\bibfield  {journal} {\bibinfo  {journal} {Proc.
  ICEM14}\ }\textbf {\bibinfo {volume} {1}},\ \bibinfo {pages} {553} (\bibinfo
  {year} {1998})}\BibitemShut {NoStop}%
\bibitem [{\citenamefont {Kathmann}\ \emph {et~al.}(2011)\citenamefont
  {Kathmann}, \citenamefont {Kuo}, \citenamefont {Mundy},\ and\ \citenamefont
  {Schenter}}]{kathmann_understanding_2011}%
  \BibitemOpen
  \bibfield  {author} {\bibinfo {author} {\bibfnamefont {S.~M.}\ \bibnamefont
  {Kathmann}}, \bibinfo {author} {\bibfnamefont {I.~W.}\ \bibnamefont {Kuo}},
  \bibinfo {author} {\bibfnamefont {C.~J.}\ \bibnamefont {Mundy}}, \ and\
  \bibinfo {author} {\bibfnamefont {G.~K.}\ \bibnamefont {Schenter}},\
  }\href@noop {} {\bibfield  {journal} {\bibinfo  {journal} {J. Phys. Chem. B}\
  }\textbf {\bibinfo {volume} {115}},\ \bibinfo {pages} {4369} (\bibinfo {year}
  {2011})}\BibitemShut {NoStop}%
\bibitem [{\citenamefont {Leung}(2010)}]{leung_surface_2010}%
  \BibitemOpen
  \bibfield  {author} {\bibinfo {author} {\bibfnamefont {K.}~\bibnamefont
  {Leung}},\ }\href@noop {} {\bibfield  {journal} {\bibinfo  {journal} {J.
  Phys. Chem. Lett.}\ }\textbf {\bibinfo {volume} {1}},\ \bibinfo {pages} {496}
  (\bibinfo {year} {2010})}\BibitemShut {NoStop}%
\bibitem [{\citenamefont {Hunt}\ and\ \citenamefont
  {Sprik}(2005)}]{sprik-sp-05}%
  \BibitemOpen
  \bibfield  {author} {\bibinfo {author} {\bibfnamefont {P.}~\bibnamefont
  {Hunt}}\ and\ \bibinfo {author} {\bibfnamefont {M.}~\bibnamefont {Sprik}},\
  }\href@noop {} {\bibfield  {journal} {\bibinfo  {journal} {ChemPhysChem}\
  }\textbf {\bibinfo {volume} {6}},\ \bibinfo {pages} {1805} (\bibinfo {year}
  {2005})}\BibitemShut {NoStop}%
\bibitem [{\citenamefont {Sokhan}\ and\ \citenamefont
  {Tildesley}(1997)}]{tildesley_wsp}%
  \BibitemOpen
  \bibfield  {author} {\bibinfo {author} {\bibfnamefont {V.~P.}\ \bibnamefont
  {Sokhan}}\ and\ \bibinfo {author} {\bibfnamefont {D.~J.}\ \bibnamefont
  {Tildesley}},\ }\href@noop {} {\bibfield  {journal} {\bibinfo  {journal}
  {Molec. Phys.}\ }\textbf {\bibinfo {volume} {92}},\ \bibinfo {pages} {625}
  (\bibinfo {year} {1997})}\BibitemShut {NoStop}%
\bibitem [{\citenamefont {Warren}\ and\ \citenamefont
  {Patel}(2007)}]{warren_hydration_2007}%
  \BibitemOpen
  \bibfield  {author} {\bibinfo {author} {\bibfnamefont {G.}~\bibnamefont
  {Warren}}\ and\ \bibinfo {author} {\bibfnamefont {S.}~\bibnamefont {Patel}},\
  }\href@noop {} {\bibfield  {journal} {\bibinfo  {journal} {J. Chem. Phys.}\
  }\textbf {\bibinfo {volume} {127}},\ \bibinfo {pages} {064509} (\bibinfo
  {year} {2007})}\BibitemShut {NoStop}%
\bibitem [{\citenamefont {Wick}\ \emph {et~al.}(2007)\citenamefont {Wick},
  \citenamefont {Kuo}, \citenamefont {Mundy},\ and\ \citenamefont
  {Dang}}]{wick_effect_2007}%
  \BibitemOpen
  \bibfield  {author} {\bibinfo {author} {\bibfnamefont {C.}~\bibnamefont
  {Wick}}, \bibinfo {author} {\bibfnamefont {I.}~\bibnamefont {Kuo}}, \bibinfo
  {author} {\bibfnamefont {C.}~\bibnamefont {Mundy}}, \ and\ \bibinfo {author}
  {\bibfnamefont {L.}~\bibnamefont {Dang}},\ }\href@noop {} {\bibfield
  {journal} {\bibinfo  {journal} {J. Chem. Theor. Comput.}\ }\textbf {\bibinfo
  {volume} {3}},\ \bibinfo {pages} {2002} (\bibinfo {year} {2007})}\BibitemShut
  {NoStop}%
\bibitem [{\citenamefont {Harder}\ and\ \citenamefont
  {Roux}(2008)}]{harder_origin_2008}%
  \BibitemOpen
  \bibfield  {author} {\bibinfo {author} {\bibfnamefont {E.}~\bibnamefont
  {Harder}}\ and\ \bibinfo {author} {\bibfnamefont {B.}~\bibnamefont {Roux}},\
  }\href@noop {} {\bibfield  {journal} {\bibinfo  {journal} {J. Chem. Phys.}\
  }\textbf {\bibinfo {volume} {129}},\ \bibinfo {pages} {234706} (\bibinfo
  {year} {2008})}\BibitemShut {NoStop}%
\bibitem [{\citenamefont {Arslanargin}\ and\ \citenamefont
  {Beck}(2012)}]{tlbsurf}%
  \BibitemOpen
  \bibfield  {author} {\bibinfo {author} {\bibfnamefont {A.}~\bibnamefont
  {Arslanargin}}\ and\ \bibinfo {author} {\bibfnamefont {T.~L.}\ \bibnamefont
  {Beck}},\ }\href@noop {} {\bibfield  {journal} {\bibinfo  {journal} {J. Chem.
  Phys.}\ }\textbf {\bibinfo {volume} {136}},\ \bibinfo {pages} {104503}
  (\bibinfo {year} {2012})}\BibitemShut {NoStop}%
\bibitem [{\citenamefont {Mundy}()}]{mundy-foot1}%
  \BibitemOpen
  \bibfield  {author} {\bibinfo {author} {\bibfnamefont {C.}~\bibnamefont
  {Mundy}},\ }\href@noop {} {\bibinfo  {journal} {{(personal communication)}}\
  }\BibitemShut {NoStop}%
\bibitem [{\citenamefont {Guggenheim}(1967)}]{guggenheim}%
  \BibitemOpen
\bibfield  {journal} {  }\bibfield  {author} {\bibinfo {author} {\bibfnamefont
  {E.~A.}\ \bibnamefont {Guggenheim}},\ }\href@noop {} {\emph {\bibinfo {title}
  {Thermodynamics: An Advanced Treatment for Chemists and Physicists}}}\
  (\bibinfo  {publisher} {North Holland, Amsterdam},\ \bibinfo {year}
  {1967})\BibitemShut {NoStop}%
\bibitem [{\citenamefont {Frumkin}(1960)}]{frumkin}%
  \BibitemOpen
  \bibfield  {author} {\bibinfo {author} {\bibfnamefont {A.~N.}\ \bibnamefont
  {Frumkin}},\ }\href@noop {} {\bibfield  {journal} {\bibinfo  {journal}
  {Electrochim. Acta}\ }\textbf {\bibinfo {volume} {2}},\ \bibinfo {pages}
  {351} (\bibinfo {year} {1960})}\BibitemShut {NoStop}%
\bibitem [{\citenamefont {Farrell}\ and\ \citenamefont
  {{McTigue}}(1982)}]{farrell}%
  \BibitemOpen
  \bibfield  {author} {\bibinfo {author} {\bibfnamefont {J.~R.}\ \bibnamefont
  {Farrell}}\ and\ \bibinfo {author} {\bibfnamefont {P.}~\bibnamefont
  {{McTigue}}},\ }\href@noop {} {\bibfield  {journal} {\bibinfo  {journal} {J.
  Electroanal. Chem.}\ }\textbf {\bibinfo {volume} {139}},\ \bibinfo {pages}
  {37} (\bibinfo {year} {1982})}\BibitemShut {NoStop}%
\bibitem [{\citenamefont {Trasatti}(1987)}]{trasatti}%
  \BibitemOpen
  \bibfield  {author} {\bibinfo {author} {\bibfnamefont {S.}~\bibnamefont
  {Trasatti}},\ }\href@noop {} {\bibfield  {journal} {\bibinfo  {journal}
  {Electrochim. Acta}\ }\textbf {\bibinfo {volume} {32}},\ \bibinfo {pages}
  {843} (\bibinfo {year} {1987})}\BibitemShut {NoStop}%
\bibitem [{\citenamefont {Kochurova}\ and\ \citenamefont
  {Rusanov}(1981)}]{rusanov}%
  \BibitemOpen
  \bibfield  {author} {\bibinfo {author} {\bibfnamefont {N.~N.}\ \bibnamefont
  {Kochurova}}\ and\ \bibinfo {author} {\bibfnamefont {A.~I.}\ \bibnamefont
  {Rusanov}},\ }\href@noop {} {\bibfield  {journal} {\bibinfo  {journal} {J.
  Colloid Interface Sci.}\ }\textbf {\bibinfo {volume} {81}},\ \bibinfo {pages}
  {297} (\bibinfo {year} {1981})}\BibitemShut {NoStop}%
\bibitem [{\citenamefont {Jarvis}\ and\ \citenamefont
  {Scheiman}(1968)}]{jarvis}%
  \BibitemOpen
  \bibfield  {author} {\bibinfo {author} {\bibfnamefont {N.~L.}\ \bibnamefont
  {Jarvis}}\ and\ \bibinfo {author} {\bibfnamefont {M.~A.}\ \bibnamefont
  {Scheiman}},\ }\href@noop {} {\bibfield  {journal} {\bibinfo  {journal} {J.
  Phys. Chem.}\ }\textbf {\bibinfo {volume} {72}},\ \bibinfo {pages} {74}
  (\bibinfo {year} {1968})}\BibitemShut {NoStop}%
\bibitem [{\citenamefont {Fawcett}(2008)}]{fawcett08}%
  \BibitemOpen
  \bibfield  {author} {\bibinfo {author} {\bibfnamefont {W.~R.}\ \bibnamefont
  {Fawcett}},\ }\href@noop {} {\bibfield  {journal} {\bibinfo  {journal}
  {Langmuir}\ }\textbf {\bibinfo {volume} {24}},\ \bibinfo {pages} {9868}
  (\bibinfo {year} {2008})}\BibitemShut {NoStop}%
\bibitem [{\citenamefont {Randles}(1977)}]{randles77}%
  \BibitemOpen
  \bibfield  {author} {\bibinfo {author} {\bibfnamefont {J.~E.~B.}\
  \bibnamefont {Randles}},\ }\href@noop {} {\bibfield  {journal} {\bibinfo
  {journal} {Phys. Chem. Liq.}\ }\textbf {\bibinfo {volume} {7}},\ \bibinfo
  {pages} {107} (\bibinfo {year} {1977})}\BibitemShut {NoStop}%
\bibitem [{\citenamefont {{de Ligny}}\ \emph {et~al.}(1969)\citenamefont {{de
  Ligny}}, \citenamefont {Alfernaar},\ and\ \citenamefont {{van der
  Veen}}}]{deligny}%
  \BibitemOpen
  \bibfield  {author} {\bibinfo {author} {\bibfnamefont {C.~L.}\ \bibnamefont
  {{de Ligny}}}, \bibinfo {author} {\bibfnamefont {M.}~\bibnamefont
  {Alfernaar}}, \ and\ \bibinfo {author} {\bibfnamefont {N.~G.}\ \bibnamefont
  {{van der Veen}}},\ }\href@noop {} {\bibfield  {journal} {\bibinfo  {journal}
  {Rec. Trav. Chim.}\ }\textbf {\bibinfo {volume} {87}},\ \bibinfo {pages}
  {585} (\bibinfo {year} {1969})}\BibitemShut {NoStop}%
\bibitem [{\citenamefont {Randles}\ and\ \citenamefont
  {Schiffrin}(1965)}]{randles65}%
  \BibitemOpen
  \bibfield  {author} {\bibinfo {author} {\bibfnamefont {J.~E.~B.}\
  \bibnamefont {Randles}}\ and\ \bibinfo {author} {\bibfnamefont {D.~J.}\
  \bibnamefont {Schiffrin}},\ }\href@noop {} {\bibfield  {journal} {\bibinfo
  {journal} {J. Elecroanal. Chem.}\ }\textbf {\bibinfo {volume} {10}},\
  \bibinfo {pages} {480} (\bibinfo {year} {1965})}\BibitemShut {NoStop}%
\bibitem [{\citenamefont {Fawcett}(2004)}]{fawcett-book}%
  \BibitemOpen
  \bibfield  {author} {\bibinfo {author} {\bibfnamefont {W.~R.}\ \bibnamefont
  {Fawcett}},\ }\href@noop {} {\emph {\bibinfo {title} {Liquids, Solutions, and
  Interfaces: from Classical Macroscopic Descriptions to Modern Microscopic
  Details}}}\ (\bibinfo  {publisher} {Oxford, New York},\ \bibinfo {year}
  {2004})\BibitemShut {NoStop}%
\bibitem [{\citenamefont {{H\"{u}nenberger}}\ and\ \citenamefont
  {Reif}(2011)}]{hunenberger-book}%
  \BibitemOpen
  \bibfield  {author} {\bibinfo {author} {\bibfnamefont {P.}~\bibnamefont
  {{H\"{u}nenberger}}}\ and\ \bibinfo {author} {\bibfnamefont {M.}~\bibnamefont
  {Reif}},\ }\href@noop {} {\emph {\bibinfo {title} {Single-Ion Solvation:
  Experimental and Theoretical Approaches to Elusive Thermodynamic
  Quantities}}}\ (\bibinfo  {publisher} {RSC Publishing, Cambridge},\ \bibinfo
  {year} {2011})\BibitemShut {NoStop}%
\bibitem [{\citenamefont {Vorobyov}\ and\ \citenamefont
  {Allen}(2010)}]{allen_SP}%
  \BibitemOpen
  \bibfield  {author} {\bibinfo {author} {\bibfnamefont {I.}~\bibnamefont
  {Vorobyov}}\ and\ \bibinfo {author} {\bibfnamefont {T.~W.}\ \bibnamefont
  {Allen}},\ }\href@noop {} {\bibfield  {journal} {\bibinfo  {journal} {J.
  Chem. Phys.}\ }\textbf {\bibinfo {volume} {132}},\ \bibinfo {pages} {185101}
  (\bibinfo {year} {2010})}\BibitemShut {NoStop}%
\bibitem [{\citenamefont {Landau}\ \emph {et~al.}(1984)\citenamefont {Landau},
  \citenamefont {Lifshitz},\ and\ \citenamefont {Pitaevskii}}]{landau8}%
  \BibitemOpen
  \bibfield  {author} {\bibinfo {author} {\bibfnamefont {L.~D.}\ \bibnamefont
  {Landau}}, \bibinfo {author} {\bibfnamefont {E.~M.}\ \bibnamefont
  {Lifshitz}}, \ and\ \bibinfo {author} {\bibfnamefont {L.~P.}\ \bibnamefont
  {Pitaevskii}},\ }\href@noop {} {\emph {\bibinfo {title} {Electrodynamics of
  Continuous Media}}}\ (\bibinfo  {publisher} {Pergamon, New York},\ \bibinfo
  {year} {1984})\BibitemShut {NoStop}%
\bibitem [{\citenamefont {Tissandier}\ \emph {et~al.}(1998)\citenamefont
  {Tissandier}, \citenamefont {Cowen}, \citenamefont {Feng}, \citenamefont
  {Gundlach}, \citenamefont {Cohen}, \citenamefont {Earhart}, \citenamefont
  {Coe},\ and\ \citenamefont {Tuttle}}]{mtiss98}%
  \BibitemOpen
  \bibfield  {author} {\bibinfo {author} {\bibfnamefont {M.~D.}\ \bibnamefont
  {Tissandier}}, \bibinfo {author} {\bibfnamefont {K.~A.}\ \bibnamefont
  {Cowen}}, \bibinfo {author} {\bibfnamefont {W.~Y.}\ \bibnamefont {Feng}},
  \bibinfo {author} {\bibfnamefont {E.}~\bibnamefont {Gundlach}}, \bibinfo
  {author} {\bibfnamefont {M.~H.}\ \bibnamefont {Cohen}}, \bibinfo {author}
  {\bibfnamefont {A.~D.}\ \bibnamefont {Earhart}}, \bibinfo {author}
  {\bibfnamefont {J.~V.}\ \bibnamefont {Coe}}, \ and\ \bibinfo {author}
  {\bibfnamefont {T.~R.}\ \bibnamefont {Tuttle}},\ }\href@noop {} {\bibfield
  {journal} {\bibinfo  {journal} {J. Phys. Chem. A}\ }\textbf {\bibinfo
  {volume} {102}},\ \bibinfo {pages} {7787} (\bibinfo {year}
  {1998})}\BibitemShut {NoStop}%
\bibitem [{\citenamefont {Marcus}(1985)}]{marcus}%
  \BibitemOpen
  \bibfield  {author} {\bibinfo {author} {\bibfnamefont {Y.}~\bibnamefont
  {Marcus}},\ }\href@noop {} {\emph {\bibinfo {title} {Ion Solvation}}}\
  (\bibinfo  {publisher} {John Wiley, New York},\ \bibinfo {year}
  {1985})\BibitemShut {NoStop}%
\bibitem [{\citenamefont {Schmid}\ \emph {et~al.}(2000)\citenamefont {Schmid},
  \citenamefont {Miah},\ and\ \citenamefont {Sapunov}}]{rschm00}%
  \BibitemOpen
  \bibfield  {author} {\bibinfo {author} {\bibfnamefont {R.}~\bibnamefont
  {Schmid}}, \bibinfo {author} {\bibfnamefont {A.~M.}\ \bibnamefont {Miah}}, \
  and\ \bibinfo {author} {\bibfnamefont {V.~N.}\ \bibnamefont {Sapunov}},\
  }\href {\doibase 10.1039/a907160a} {\bibfield  {journal} {\bibinfo  {journal}
  {Phys. Chem. Chem. Phys.}\ }\textbf {\bibinfo {volume} {2}},\ \bibinfo
  {pages} {97} (\bibinfo {year} {2000})}\BibitemShut {NoStop}%
\bibitem [{\citenamefont {Hummer}\ \emph {et~al.}(1996)\citenamefont {Hummer},
  \citenamefont {Pratt},\ and\ \citenamefont {Garcia}}]{ghumm96}%
  \BibitemOpen
  \bibfield  {author} {\bibinfo {author} {\bibfnamefont {G.}~\bibnamefont
  {Hummer}}, \bibinfo {author} {\bibfnamefont {L.~R.}\ \bibnamefont {Pratt}}, \
  and\ \bibinfo {author} {\bibfnamefont {A.~E.}\ \bibnamefont {Garcia}},\
  }\href@noop {} {\bibfield  {journal} {\bibinfo  {journal} {J. Phys. Chem.}\
  }\textbf {\bibinfo {volume} {100}},\ \bibinfo {pages} {1206} (\bibinfo {year}
  {1996})}\BibitemShut {NoStop}%
\bibitem [{\citenamefont {Ashbaugh}(2000)}]{hank-phi}%
  \BibitemOpen
  \bibfield  {author} {\bibinfo {author} {\bibfnamefont {H.~S.}\ \bibnamefont
  {Ashbaugh}},\ }\href@noop {} {\bibfield  {journal} {\bibinfo  {journal} {J.
  Phys. Chem. B}\ }\textbf {\bibinfo {volume} {104}},\ \bibinfo {pages} {7235}
  (\bibinfo {year} {2000})}\BibitemShut {NoStop}%
\bibitem [{\citenamefont {Rajamani}\ \emph {et~al.}(2004)\citenamefont
  {Rajamani}, \citenamefont {Ghosh},\ and\ \citenamefont {Garde}}]{garde-phi}%
  \BibitemOpen
  \bibfield  {author} {\bibinfo {author} {\bibfnamefont {S.}~\bibnamefont
  {Rajamani}}, \bibinfo {author} {\bibfnamefont {T.}~\bibnamefont {Ghosh}}, \
  and\ \bibinfo {author} {\bibfnamefont {S.}~\bibnamefont {Garde}},\
  }\href@noop {} {\bibfield  {journal} {\bibinfo  {journal} {J. Chem. Phys.}\
  }\textbf {\bibinfo {volume} {120}},\ \bibinfo {pages} {4457} (\bibinfo {year}
  {2004})}\BibitemShut {NoStop}%
\bibitem [{\citenamefont {Rogers}\ and\ \citenamefont {Beck}(2010)}]{droge10}%
  \BibitemOpen
  \bibfield  {author} {\bibinfo {author} {\bibfnamefont {D.~M.}\ \bibnamefont
  {Rogers}}\ and\ \bibinfo {author} {\bibfnamefont {T.~L.}\ \bibnamefont
  {Beck}},\ }\href@noop {} {\bibfield  {journal} {\bibinfo  {journal} {J. Chem.
  Phys.}\ }\textbf {\bibinfo {volume} {132}},\ \bibinfo {pages} {014505}
  (\bibinfo {year} {2010})}\BibitemShut {NoStop}%
\bibitem [{\citenamefont {Baer}\ \emph {et~al.}(2012)\citenamefont {Baer},
  \citenamefont {Stern}, \citenamefont {Levin}, \citenamefont {Tobias},\ and\
  \citenamefont {Mundy}}]{mundy-SP-12}%
  \BibitemOpen
  \bibfield  {author} {\bibinfo {author} {\bibfnamefont {M.~D.}\ \bibnamefont
  {Baer}}, \bibinfo {author} {\bibfnamefont {A.~C.}\ \bibnamefont {Stern}},
  \bibinfo {author} {\bibfnamefont {Y.}~\bibnamefont {Levin}}, \bibinfo
  {author} {\bibfnamefont {D.~J.}\ \bibnamefont {Tobias}}, \ and\ \bibinfo
  {author} {\bibfnamefont {C.~J.}\ \bibnamefont {Mundy}},\ }\href@noop {}
  {\bibfield  {journal} {\bibinfo  {journal} {J. Phys. Chem. Lett.}\ }\textbf
  {\bibinfo {volume} {3}},\ \bibinfo {pages} {1565} (\bibinfo {year}
  {2012})}\BibitemShut {NoStop}%
\bibitem [{\citenamefont {Otten}\ \emph {et~al.}(2012)\citenamefont {Otten},
  \citenamefont {Shaffer}, \citenamefont {Geissler},\ and\ \citenamefont
  {Saykally}}]{saykally12}%
  \BibitemOpen
  \bibfield  {author} {\bibinfo {author} {\bibfnamefont {D.~E.}\ \bibnamefont
  {Otten}}, \bibinfo {author} {\bibfnamefont {P.~R.}\ \bibnamefont {Shaffer}},
  \bibinfo {author} {\bibfnamefont {P.~L.}\ \bibnamefont {Geissler}}, \ and\
  \bibinfo {author} {\bibfnamefont {R.~J.}\ \bibnamefont {Saykally}},\
  }\href@noop {} {\bibfield  {journal} {\bibinfo  {journal} {Proc. Natl. Acad.
  Sci. USA}\ }\textbf {\bibinfo {volume} {109}},\ \bibinfo {pages} {701}
  (\bibinfo {year} {2012})}\BibitemShut {NoStop}%
\bibitem [{\citenamefont {Pratt}(1992)}]{pratt_contact_1992}%
  \BibitemOpen
  \bibfield  {author} {\bibinfo {author} {\bibfnamefont {L.~R.}\ \bibnamefont
  {Pratt}},\ }\href@noop {} {\bibfield  {journal} {\bibinfo  {journal} {Journal
  of Physical Chemistry}\ }\textbf {\bibinfo {volume} {96}},\ \bibinfo {pages}
  {25} (\bibinfo {year} {1992})}\BibitemShut {NoStop}%
\bibitem [{\citenamefont {Wilson}\ \emph {et~al.}(1988)\citenamefont {Wilson},
  \citenamefont {Pohorille},\ and\ \citenamefont {Pratt}}]{mwils88}%
  \BibitemOpen
  \bibfield  {author} {\bibinfo {author} {\bibfnamefont {M.~A.}\ \bibnamefont
  {Wilson}}, \bibinfo {author} {\bibfnamefont {A.}~\bibnamefont {Pohorille}}, \
  and\ \bibinfo {author} {\bibfnamefont {L.~R.}\ \bibnamefont {Pratt}},\
  }\href@noop {} {\bibfield  {journal} {\bibinfo  {journal} {J. Chem. Phys.}\
  }\textbf {\bibinfo {volume} {88}},\ \bibinfo {pages} {3281} (\bibinfo {year}
  {1988})}\BibitemShut {NoStop}%
\bibitem [{\citenamefont {Wilson}\ \emph {et~al.}(1989)\citenamefont {Wilson},
  \citenamefont {Pohorille},\ and\ \citenamefont {Pratt}}]{lrp_SP_89}%
  \BibitemOpen
  \bibfield  {author} {\bibinfo {author} {\bibfnamefont {M.~A.}\ \bibnamefont
  {Wilson}}, \bibinfo {author} {\bibfnamefont {A.}~\bibnamefont {Pohorille}}, \
  and\ \bibinfo {author} {\bibfnamefont {L.~R.}\ \bibnamefont {Pratt}},\
  }\href@noop {} {\bibfield  {journal} {\bibinfo  {journal} {J. Chem. Phys.}\
  }\textbf {\bibinfo {volume} {90}},\ \bibinfo {pages} {5211} (\bibinfo {year}
  {1989})}\BibitemShut {NoStop}%
\bibitem [{\citenamefont {Jackson}(1998)}]{jackson}%
  \BibitemOpen
  \bibfield  {author} {\bibinfo {author} {\bibfnamefont {J.~D.}\ \bibnamefont
  {Jackson}},\ }\href@noop {} {\emph {\bibinfo {title} {Classical
  Electrodynamics}}},\ \bibinfo {edition} {3rd}\ ed.\ (\bibinfo  {publisher}
  {John Wiley, New York},\ \bibinfo {year} {1998})\BibitemShut {NoStop}%
\bibitem [{\citenamefont {Schmickler}(1996)}]{schmickler}%
  \BibitemOpen
  \bibfield  {author} {\bibinfo {author} {\bibfnamefont {W.}~\bibnamefont
  {Schmickler}},\ }\href@noop {} {\emph {\bibinfo {title} {Interfacial
  Electrochemistry}}}\ (\bibinfo  {publisher} {Oxford University Press,
  Oxford},\ \bibinfo {year} {1996})\BibitemShut {NoStop}%
\bibitem [{\citenamefont {Hamann}\ \emph {et~al.}(1998)\citenamefont {Hamann},
  \citenamefont {Hamnett},\ and\ \citenamefont {Vielstich}}]{hamann}%
  \BibitemOpen
  \bibfield  {author} {\bibinfo {author} {\bibfnamefont {C.~H.}\ \bibnamefont
  {Hamann}}, \bibinfo {author} {\bibfnamefont {A.}~\bibnamefont {Hamnett}}, \
  and\ \bibinfo {author} {\bibfnamefont {W.}~\bibnamefont {Vielstich}},\
  }\href@noop {} {\emph {\bibinfo {title} {Electrochemistry}}}\ (\bibinfo
  {publisher} {Wiley VCH, Weinheim},\ \bibinfo {year} {1998})\BibitemShut
  {NoStop}%
\bibitem [{\citenamefont {Lamoureux}\ and\ \citenamefont
  {Roux}(2006{\natexlab{a}})}]{lamoureux_absolute_2006}%
  \BibitemOpen
  \bibfield  {author} {\bibinfo {author} {\bibfnamefont {G.}~\bibnamefont
  {Lamoureux}}\ and\ \bibinfo {author} {\bibfnamefont {B.}~\bibnamefont
  {Roux}},\ }\href@noop {} {\bibfield  {journal} {\bibinfo  {journal} {J. Phys.
  Chem. B}\ }\textbf {\bibinfo {volume} {110}},\ \bibinfo {pages} {3308}
  (\bibinfo {year} {2006}{\natexlab{a}})}\BibitemShut {NoStop}%
\bibitem [{\citenamefont {Leung}\ \emph {et~al.}(2009)\citenamefont {Leung},
  \citenamefont {Rempe},\ and\ \citenamefont {von Lilienfeld}}]{leung_ab_2009}%
  \BibitemOpen
  \bibfield  {author} {\bibinfo {author} {\bibfnamefont {K.}~\bibnamefont
  {Leung}}, \bibinfo {author} {\bibfnamefont {S.}~\bibnamefont {Rempe}}, \ and\
  \bibinfo {author} {\bibfnamefont {O.}~\bibnamefont {von Lilienfeld}},\
  }\href@noop {} {\bibfield  {journal} {\bibinfo  {journal} {J. Chem. Phys.}\
  }\textbf {\bibinfo {volume} {130}} (\bibinfo {year} {2009})}\BibitemShut
  {NoStop}%
\bibitem [{\citenamefont {Rempe}\ and\ \citenamefont
  {Leung}(2010)}]{leung_ab_2009_rep}%
  \BibitemOpen
  \bibfield  {author} {\bibinfo {author} {\bibfnamefont {S.~B.}\ \bibnamefont
  {Rempe}}\ and\ \bibinfo {author} {\bibfnamefont {K.}~\bibnamefont {Leung}},\
  }\href@noop {} {\bibfield  {journal} {\bibinfo  {journal} {J. Chem. Phys.}\
  }\textbf {\bibinfo {volume} {133}},\ \bibinfo {pages} {047103} (\bibinfo
  {year} {2010})}\BibitemShut {NoStop}%
\bibitem [{\citenamefont {Latimer}\ \emph {et~al.}(1939)\citenamefont
  {Latimer}, \citenamefont {Pitzer},\ and\ \citenamefont
  {Slansky}}]{latimer_39}%
  \BibitemOpen
  \bibfield  {author} {\bibinfo {author} {\bibfnamefont {W.~M.}\ \bibnamefont
  {Latimer}}, \bibinfo {author} {\bibfnamefont {K.~S.}\ \bibnamefont {Pitzer}},
  \ and\ \bibinfo {author} {\bibfnamefont {C.~M.}\ \bibnamefont {Slansky}},\
  }\href@noop {} {\bibfield  {journal} {\bibinfo  {journal} {J. Chem. Phys.}\
  }\textbf {\bibinfo {volume} {7}},\ \bibinfo {pages} {108} (\bibinfo {year}
  {1939})}\BibitemShut {NoStop}%
\bibitem [{\citenamefont {Ashbaugh}\ and\ \citenamefont
  {Asthagiri}(2008)}]{ashbaugh_single_2008}%
  \BibitemOpen
  \bibfield  {author} {\bibinfo {author} {\bibfnamefont {H.}~\bibnamefont
  {Ashbaugh}}\ and\ \bibinfo {author} {\bibfnamefont {D.}~\bibnamefont
  {Asthagiri}},\ }\href@noop {} {\bibfield  {journal} {\bibinfo  {journal} {J.
  Chem. Phys.}\ }\textbf {\bibinfo {volume} {129}},\ \bibinfo {pages} {204501}
  (\bibinfo {year} {2008})}\BibitemShut {NoStop}%
\bibitem [{\citenamefont {Widom}(1963)}]{bwido63}%
  \BibitemOpen
  \bibfield  {author} {\bibinfo {author} {\bibfnamefont {B.}~\bibnamefont
  {Widom}},\ }\href@noop {} {\bibfield  {journal} {\bibinfo  {journal} {J.
  Chem. Phys.}\ }\textbf {\bibinfo {volume} {39}},\ \bibinfo {pages} {2808}
  (\bibinfo {year} {1963})}\BibitemShut {NoStop}%
\bibitem [{\citenamefont {Beck}\ \emph {et~al.}(2006)\citenamefont {Beck},
  \citenamefont {Paulaitis},\ and\ \citenamefont {Pratt}}]{ourbook}%
  \BibitemOpen
  \bibfield  {author} {\bibinfo {author} {\bibfnamefont {T.~L.}\ \bibnamefont
  {Beck}}, \bibinfo {author} {\bibfnamefont {M.~E.}\ \bibnamefont {Paulaitis}},
  \ and\ \bibinfo {author} {\bibfnamefont {L.~R.}\ \bibnamefont {Pratt}},\
  }\href@noop {} {\emph {\bibinfo {title} {The Potential Distribution Theorem
  and Models of Molecular Solutions}}}\ (\bibinfo  {publisher} {Cambridge, New
  York},\ \bibinfo {year} {2006})\BibitemShut {NoStop}%
\bibitem [{\citenamefont {Beck}(2011{\natexlab{a}})}]{tlbent11}%
  \BibitemOpen
  \bibfield  {author} {\bibinfo {author} {\bibfnamefont {T.~L.}\ \bibnamefont
  {Beck}},\ }\href@noop {} {\bibfield  {journal} {\bibinfo  {journal} {J. Phys.
  Chem. B}\ }\textbf {\bibinfo {volume} {115}},\ \bibinfo {pages} {9776}
  (\bibinfo {year} {2011}{\natexlab{a}})}\BibitemShut {NoStop}%
\bibitem [{\citenamefont {Beck}(2011{\natexlab{b}})}]{tlbloc11}%
  \BibitemOpen
  \bibfield  {author} {\bibinfo {author} {\bibfnamefont {T.~L.}\ \bibnamefont
  {Beck}},\ }\href@noop {} {\bibfield  {journal} {\bibinfo  {journal} {J. Stat.
  Phys.}\ }\textbf {\bibinfo {volume} {145}},\ \bibinfo {pages} {335} (\bibinfo
  {year} {2011}{\natexlab{b}})}\BibitemShut {NoStop}%
\bibitem [{\citenamefont {Pratt}\ and\ \citenamefont
  {LaViolette}(1998)}]{lrprat98}%
  \BibitemOpen
  \bibfield  {author} {\bibinfo {author} {\bibfnamefont {L.~R.}\ \bibnamefont
  {Pratt}}\ and\ \bibinfo {author} {\bibfnamefont {R.~A.}\ \bibnamefont
  {LaViolette}},\ }\href@noop {} {\bibfield  {journal} {\bibinfo  {journal}
  {Mol. Phys.}\ }\textbf {\bibinfo {volume} {94}},\ \bibinfo {pages} {909}
  (\bibinfo {year} {1998})}\BibitemShut {NoStop}%
\bibitem [{\citenamefont {Pratt}\ and\ \citenamefont
  {Rempe}(1999)}]{lrprat991}%
  \BibitemOpen
  \bibfield  {author} {\bibinfo {author} {\bibfnamefont {L.~R.}\ \bibnamefont
  {Pratt}}\ and\ \bibinfo {author} {\bibfnamefont {S.~B.}\ \bibnamefont
  {Rempe}},\ }in\ \href@noop {} {\emph {\bibinfo {booktitle} {Simulation and
  Theory of Electrostatic Interactions in Solution}}},\ \bibinfo {editor}
  {edited by\ \bibinfo {editor} {\bibfnamefont {G.}~\bibnamefont {Hummer}}\
  and\ \bibinfo {editor} {\bibfnamefont {L.~R.}\ \bibnamefont {Pratt}}}\
  (\bibinfo  {publisher} {AIP Press, New York},\ \bibinfo {year} {1999})\ pp.\
  \bibinfo {pages} {172--201}\BibitemShut {NoStop}%
\bibitem [{\citenamefont {Asthagiri}\ \emph {et~al.}(2010)\citenamefont
  {Asthagiri}, \citenamefont {Dixit}, \citenamefont {Merchant}, \citenamefont
  {Paulaitis}, \citenamefont {Pratt}, \citenamefont {Rempe},\ and\
  \citenamefont {Varma}}]{lrp_coordination_10}%
  \BibitemOpen
  \bibfield  {author} {\bibinfo {author} {\bibfnamefont {D.}~\bibnamefont
  {Asthagiri}}, \bibinfo {author} {\bibfnamefont {P.~D.}\ \bibnamefont
  {Dixit}}, \bibinfo {author} {\bibfnamefont {S.}~\bibnamefont {Merchant}},
  \bibinfo {author} {\bibfnamefont {M.~E.}\ \bibnamefont {Paulaitis}}, \bibinfo
  {author} {\bibfnamefont {L.~R.}\ \bibnamefont {Pratt}}, \bibinfo {author}
  {\bibfnamefont {S.~B.}\ \bibnamefont {Rempe}}, \ and\ \bibinfo {author}
  {\bibfnamefont {S.}~\bibnamefont {Varma}},\ }\href@noop {} {\bibfield
  {journal} {\bibinfo  {journal} {Chem. Phys. Lett.}\ }\textbf {\bibinfo
  {volume} {485}},\ \bibinfo {pages} {1} (\bibinfo {year} {2010})}\BibitemShut
  {NoStop}%
\bibitem [{\citenamefont {Rodgers}\ and\ \citenamefont
  {Weeks}(2008{\natexlab{a}})}]{rodgers_local_2008}%
  \BibitemOpen
  \bibfield  {author} {\bibinfo {author} {\bibfnamefont {J.}~\bibnamefont
  {Rodgers}}\ and\ \bibinfo {author} {\bibfnamefont {J.}~\bibnamefont
  {Weeks}},\ }\href@noop {} {\bibfield  {journal} {\bibinfo  {journal} {J.
  Phys. Cond. Matter}\ }\textbf {\bibinfo {volume} {20}},\ \bibinfo {pages}
  {494206} (\bibinfo {year} {2008}{\natexlab{a}})}\BibitemShut {NoStop}%
\bibitem [{\citenamefont {Chempath}\ \emph {et~al.}(2009)\citenamefont
  {Chempath}, \citenamefont {Pratt},\ and\ \citenamefont
  {Paulaitis}}]{chempath_quasichemical_2009}%
  \BibitemOpen
  \bibfield  {author} {\bibinfo {author} {\bibfnamefont {S.}~\bibnamefont
  {Chempath}}, \bibinfo {author} {\bibfnamefont {L.}~\bibnamefont {Pratt}}, \
  and\ \bibinfo {author} {\bibfnamefont {M.}~\bibnamefont {Paulaitis}},\
  }\href@noop {} {\bibfield  {journal} {\bibinfo  {journal} {J. Chem. Phys.}\
  }\textbf {\bibinfo {volume} {130}},\ \bibinfo {pages} {054113} (\bibinfo
  {year} {2009})}\BibitemShut {NoStop}%
\bibitem [{\citenamefont {Weber}\ \emph {et~al.}(2011)\citenamefont {Weber},
  \citenamefont {Merchant},\ and\ \citenamefont
  {Asthagiri}}]{dilip-regularize}%
  \BibitemOpen
  \bibfield  {author} {\bibinfo {author} {\bibfnamefont {V.}~\bibnamefont
  {Weber}}, \bibinfo {author} {\bibfnamefont {S.}~\bibnamefont {Merchant}}, \
  and\ \bibinfo {author} {\bibfnamefont {D.}~\bibnamefont {Asthagiri}},\
  }\href@noop {} {\bibfield  {journal} {\bibinfo  {journal} {J. Chem. Phys.}\
  }\textbf {\bibinfo {volume} {135}},\ \bibinfo {pages} {181101} (\bibinfo
  {year} {2011})}\BibitemShut {NoStop}%
\bibitem [{\citenamefont {Rodgers}\ and\ \citenamefont
  {Weeks}(2008{\natexlab{b}})}]{rodgers_interplay_2008}%
  \BibitemOpen
  \bibfield  {author} {\bibinfo {author} {\bibfnamefont {J.}~\bibnamefont
  {Rodgers}}\ and\ \bibinfo {author} {\bibfnamefont {J.}~\bibnamefont
  {Weeks}},\ }\href {\doibase 10.1073/pnas.0807623105} {\bibfield  {journal}
  {\bibinfo  {journal} {Proc. Natl. Acad. Sci. USA}\ }\textbf {\bibinfo
  {volume} {105}},\ \bibinfo {pages} {19136} (\bibinfo {year}
  {2008}{\natexlab{b}})}\BibitemShut {NoStop}%
\bibitem [{\citenamefont {Rodgers}\ and\ \citenamefont
  {Weeks}(2009)}]{rodgers_accurate_2009}%
  \BibitemOpen
  \bibfield  {author} {\bibinfo {author} {\bibfnamefont {J.}~\bibnamefont
  {Rodgers}}\ and\ \bibinfo {author} {\bibfnamefont {J.}~\bibnamefont
  {Weeks}},\ }\href {\doibase 10.1063/1.3276729} {\bibfield  {journal}
  {\bibinfo  {journal} {J. Chem. Phys.}\ }\textbf {\bibinfo {volume} {131}},\
  \bibinfo {pages} {244108} (\bibinfo {year} {2009})}\BibitemShut {NoStop}%
\bibitem [{\citenamefont {Horinek}\ \emph
  {et~al.}(2009{\natexlab{a}})\citenamefont {Horinek}, \citenamefont
  {Mamatkulov},\ and\ \citenamefont {Netz}}]{horinek_rational_2009}%
  \BibitemOpen
  \bibfield  {author} {\bibinfo {author} {\bibfnamefont {D.}~\bibnamefont
  {Horinek}}, \bibinfo {author} {\bibfnamefont {S.}~\bibnamefont {Mamatkulov}},
  \ and\ \bibinfo {author} {\bibfnamefont {R.}~\bibnamefont {Netz}},\
  }\href@noop {} {\bibfield  {journal} {\bibinfo  {journal} {J. Chem. Phys.}\
  }\textbf {\bibinfo {volume} {130}},\ \bibinfo {pages} {124507} (\bibinfo
  {year} {2009}{\natexlab{a}})}\BibitemShut {NoStop}%
\bibitem [{\citenamefont {Asthagiri}\ and\ \citenamefont
  {Pratt}(2003)}]{asthagiri_quasi-chemical_2003}%
  \BibitemOpen
  \bibfield  {author} {\bibinfo {author} {\bibfnamefont {D.}~\bibnamefont
  {Asthagiri}}\ and\ \bibinfo {author} {\bibfnamefont {L.}~\bibnamefont
  {Pratt}},\ }\href {\doibase 10.1016/S0009-2614(03)00227-6} {\bibfield
  {journal} {\bibinfo  {journal} {Chem. Phys. Lett.}\ }\textbf {\bibinfo
  {volume} {371}},\ \bibinfo {pages} {613} (\bibinfo {year}
  {2003})}\BibitemShut {NoStop}%
\bibitem [{\citenamefont {Kelly}\ \emph {et~al.}(2006)\citenamefont {Kelly},
  \citenamefont {Cramer},\ and\ \citenamefont {Truhlar}}]{kelly_aqueous_2006}%
  \BibitemOpen
  \bibfield  {author} {\bibinfo {author} {\bibfnamefont {C.~P.}\ \bibnamefont
  {Kelly}}, \bibinfo {author} {\bibfnamefont {C.~J.}\ \bibnamefont {Cramer}}, \
  and\ \bibinfo {author} {\bibfnamefont {D.~G.}\ \bibnamefont {Truhlar}},\
  }\href@noop {} {\bibfield  {journal} {\bibinfo  {journal} {J. Phys. Chem. B}\
  }\textbf {\bibinfo {volume} {110}},\ \bibinfo {pages} {16066} (\bibinfo
  {year} {2006})}\BibitemShut {NoStop}%
\bibitem [{\citenamefont {Asthagiri}\ \emph {et~al.}(2003)\citenamefont
  {Asthagiri}, \citenamefont {Pratt},\ and\ \citenamefont
  {Ashbaugh}}]{asthagiri_absolute_2003}%
  \BibitemOpen
  \bibfield  {author} {\bibinfo {author} {\bibfnamefont {D.}~\bibnamefont
  {Asthagiri}}, \bibinfo {author} {\bibfnamefont {L.}~\bibnamefont {Pratt}}, \
  and\ \bibinfo {author} {\bibfnamefont {H.}~\bibnamefont {Ashbaugh}},\ }\href
  {\doibase 10.1063/1.1587122} {\bibfield  {journal} {\bibinfo  {journal} {J.
  Chem. Phys.}\ }\textbf {\bibinfo {volume} {119}},\ \bibinfo {pages} {2702}
  (\bibinfo {year} {2003})}\BibitemShut {NoStop}%
\bibitem [{\citenamefont {Ben-Amotz}\ \emph {et~al.}(2005)\citenamefont
  {Ben-Amotz}, \citenamefont {Raineri},\ and\ \citenamefont
  {Stell}}]{ben-amotz-theory1}%
  \BibitemOpen
  \bibfield  {author} {\bibinfo {author} {\bibfnamefont {D.}~\bibnamefont
  {Ben-Amotz}}, \bibinfo {author} {\bibfnamefont {F.~O.}\ \bibnamefont
  {Raineri}}, \ and\ \bibinfo {author} {\bibfnamefont {G.}~\bibnamefont
  {Stell}},\ }\href@noop {} {\bibfield  {journal} {\bibinfo  {journal} {J.
  Phys. Chem. B}\ }\textbf {\bibinfo {volume} {109}},\ \bibinfo {pages} {6866}
  (\bibinfo {year} {2005})}\BibitemShut {NoStop}%
\bibitem [{\citenamefont {Yagasaki}\ \emph {et~al.}(2010)\citenamefont
  {Yagasaki}, \citenamefont {Saito},\ and\ \citenamefont
  {Ohmine}}]{yagasaki_effects_2010}%
  \BibitemOpen
  \bibfield  {author} {\bibinfo {author} {\bibfnamefont {T.}~\bibnamefont
  {Yagasaki}}, \bibinfo {author} {\bibfnamefont {S.}~\bibnamefont {Saito}}, \
  and\ \bibinfo {author} {\bibfnamefont {I.}~\bibnamefont {Ohmine}},\
  }\href@noop {} {\bibfield  {journal} {\bibinfo  {journal} {J. Phys. Chem. A}\
  }\textbf {\bibinfo {volume} {114}},\ \bibinfo {pages} {12573} (\bibinfo
  {year} {2010})}\BibitemShut {NoStop}%
\bibitem [{\citenamefont {Caleman}\ \emph {et~al.}(2011)\citenamefont
  {Caleman}, \citenamefont {Hub}, \citenamefont {{van Maaren}},\ and\
  \citenamefont {{van der Spoel}}}]{swedes}%
  \BibitemOpen
  \bibfield  {author} {\bibinfo {author} {\bibfnamefont {C.}~\bibnamefont
  {Caleman}}, \bibinfo {author} {\bibfnamefont {J.~S.}\ \bibnamefont {Hub}},
  \bibinfo {author} {\bibfnamefont {P.~J.}\ \bibnamefont {{van Maaren}}}, \
  and\ \bibinfo {author} {\bibfnamefont {D.}~\bibnamefont {{van der Spoel}}},\
  }\href@noop {} {\bibfield  {journal} {\bibinfo  {journal} {Proc. Natl. Acad.
  Sci. USA}\ }\textbf {\bibinfo {volume} {108}},\ \bibinfo {pages} {6838}
  (\bibinfo {year} {2011})}\BibitemShut {NoStop}%
\bibitem [{\citenamefont {Ponder}(2012)}]{tinker}%
  \BibitemOpen
  \bibfield  {author} {\bibinfo {author} {\bibfnamefont {J.~W.}\ \bibnamefont
  {Ponder}},\ }\href@noop {} {\bibfield  {journal} {\bibinfo  {journal}
  {{TINKER}: Software Tools for Molecular Design, Version 6.1; Saint Louis,
  MO}\ } (\bibinfo {year} {2012})}\BibitemShut {NoStop}%
\bibitem [{\citenamefont {Neese}(2012)}]{ORCA}%
  \BibitemOpen
  \bibfield  {author} {\bibinfo {author} {\bibfnamefont {F.}~\bibnamefont
  {Neese}},\ }\href@noop {} {\bibfield  {journal} {\bibinfo  {journal} {ORCA
  electronic structure code}\ ,\ \bibinfo {pages}
  {http://www.mpibac.mpg.de/bac/logins/neese/description.php}} (\bibinfo {year}
  {2012})}\BibitemShut {NoStop}%
\bibitem [{\citenamefont {Neese}\ \emph {et~al.}(2011)\citenamefont {Neese},
  \citenamefont {Liakos},\ and\ \citenamefont {Ye}}]{neese}%
  \BibitemOpen
  \bibfield  {author} {\bibinfo {author} {\bibfnamefont {F.}~\bibnamefont
  {Neese}}, \bibinfo {author} {\bibfnamefont {D.~G.}\ \bibnamefont {Liakos}}, \
  and\ \bibinfo {author} {\bibfnamefont {S.~F.}\ \bibnamefont {Ye}},\
  }\href@noop {} {\bibfield  {journal} {\bibinfo  {journal} {J. Biol. Inorg.
  Chem.}\ }\textbf {\bibinfo {volume} {16}},\ \bibinfo {pages} {821} (\bibinfo
  {year} {2011})}\BibitemShut {NoStop}%
\bibitem [{\citenamefont {Horinek}\ \emph
  {et~al.}(2009{\natexlab{b}})\citenamefont {Horinek}, \citenamefont {Herz},
  \citenamefont {Vrbka}, \citenamefont {Sedlmeier}, \citenamefont
  {Mamatkulov},\ and\ \citenamefont {Netz}}]{horinek_specific_2009}%
  \BibitemOpen
  \bibfield  {author} {\bibinfo {author} {\bibfnamefont {D.}~\bibnamefont
  {Horinek}}, \bibinfo {author} {\bibfnamefont {A.}~\bibnamefont {Herz}},
  \bibinfo {author} {\bibfnamefont {L.}~\bibnamefont {Vrbka}}, \bibinfo
  {author} {\bibfnamefont {F.}~\bibnamefont {Sedlmeier}}, \bibinfo {author}
  {\bibfnamefont {S.~I.}\ \bibnamefont {Mamatkulov}}, \ and\ \bibinfo {author}
  {\bibfnamefont {R.~R.}\ \bibnamefont {Netz}},\ }\href@noop {} {\bibfield
  {journal} {\bibinfo  {journal} {Chem. Phys. Lett.}\ }\textbf {\bibinfo
  {volume} {479}},\ \bibinfo {pages} {173} (\bibinfo {year}
  {2009}{\natexlab{b}})}\BibitemShut {NoStop}%
\bibitem [{\citenamefont {Fernandez}\ \emph {et~al.}(2006)\citenamefont
  {Fernandez}, \citenamefont {Abascal},\ and\ \citenamefont
  {Vega}}]{spce-freeze}%
  \BibitemOpen
  \bibfield  {author} {\bibinfo {author} {\bibfnamefont {R.~G.}\ \bibnamefont
  {Fernandez}}, \bibinfo {author} {\bibfnamefont {J.~L.}\ \bibnamefont
  {Abascal}}, \ and\ \bibinfo {author} {\bibfnamefont {C.}~\bibnamefont
  {Vega}},\ }\href@noop {} {\bibfield  {journal} {\bibinfo  {journal} {J. Chem.
  Phys.}\ }\textbf {\bibinfo {volume} {124}},\ \bibinfo {pages} {144506}
  (\bibinfo {year} {2006})}\BibitemShut {NoStop}%
\bibitem [{\citenamefont {Fan}\ \emph {et~al.}(2009)\citenamefont {Fan},
  \citenamefont {Chen}, \citenamefont {Yang}, \citenamefont {Cremer},\ and\
  \citenamefont {Gao}}]{cremer-spcevsexpt}%
  \BibitemOpen
  \bibfield  {author} {\bibinfo {author} {\bibfnamefont {Y.}~\bibnamefont
  {Fan}}, \bibinfo {author} {\bibfnamefont {X.}~\bibnamefont {Chen}}, \bibinfo
  {author} {\bibfnamefont {L.}~\bibnamefont {Yang}}, \bibinfo {author}
  {\bibfnamefont {P.~S.}\ \bibnamefont {Cremer}}, \ and\ \bibinfo {author}
  {\bibfnamefont {Y.~Q.}\ \bibnamefont {Gao}},\ }\href@noop {} {\bibfield
  {journal} {\bibinfo  {journal} {J. Phys. Chem. B}\ }\textbf {\bibinfo
  {volume} {113}},\ \bibinfo {pages} {11672} (\bibinfo {year}
  {2009})}\BibitemShut {NoStop}%
\bibitem [{\citenamefont {Kuo}\ \emph {et~al.}(2006)\citenamefont {Kuo},
  \citenamefont {Mundy}, \citenamefont {Eggimann}, \citenamefont {{McGrath}},
  \citenamefont {Siepmann}, \citenamefont {Chen}, \citenamefont {Vieceli},\
  and\ \citenamefont {Tobias}}]{mundy-aimdvspc}%
  \BibitemOpen
  \bibfield  {author} {\bibinfo {author} {\bibfnamefont {I.}~\bibnamefont
  {Kuo}}, \bibinfo {author} {\bibfnamefont {C.~J.}\ \bibnamefont {Mundy}},
  \bibinfo {author} {\bibfnamefont {B.~L.}\ \bibnamefont {Eggimann}}, \bibinfo
  {author} {\bibfnamefont {M.~J.}\ \bibnamefont {{McGrath}}}, \bibinfo {author}
  {\bibfnamefont {J.~I.}\ \bibnamefont {Siepmann}}, \bibinfo {author}
  {\bibfnamefont {B.}~\bibnamefont {Chen}}, \bibinfo {author} {\bibfnamefont
  {J.}~\bibnamefont {Vieceli}}, \ and\ \bibinfo {author} {\bibfnamefont
  {D.~J.}\ \bibnamefont {Tobias}},\ }\href@noop {} {\bibfield  {journal}
  {\bibinfo  {journal} {J. Phys. Chem. B}\ }\textbf {\bibinfo {volume} {110}},\
  \bibinfo {pages} {3738} (\bibinfo {year} {2006})}\BibitemShut {NoStop}%
\bibitem [{\citenamefont {K\"{u}hne}\ \emph {et~al.}(2011)\citenamefont
  {K\"{u}hne}, \citenamefont {Pascal}, \citenamefont {Kaxiras},\ and\
  \citenamefont {Jung}}]{kaxiras-aimdvspc}%
  \BibitemOpen
  \bibfield  {author} {\bibinfo {author} {\bibfnamefont {T.~D.}\ \bibnamefont
  {K\"{u}hne}}, \bibinfo {author} {\bibfnamefont {T.~A.}\ \bibnamefont
  {Pascal}}, \bibinfo {author} {\bibfnamefont {E.}~\bibnamefont {Kaxiras}}, \
  and\ \bibinfo {author} {\bibfnamefont {Y.}~\bibnamefont {Jung}},\ }\href@noop
  {} {\bibfield  {journal} {\bibinfo  {journal} {J. Phys. Chem. Lett.}\
  }\textbf {\bibinfo {volume} {2}},\ \bibinfo {pages} {105} (\bibinfo {year}
  {2011})}\BibitemShut {NoStop}%
\bibitem [{\citenamefont {Gu{\'a}rdia}\ \emph {et~al.}(2009)\citenamefont
  {Gu{\'a}rdia}, \citenamefont {Skarmoutsos},\ and\ \citenamefont
  {Masia}}]{eguar09}%
  \BibitemOpen
  \bibfield  {author} {\bibinfo {author} {\bibfnamefont {E.}~\bibnamefont
  {Gu{\'a}rdia}}, \bibinfo {author} {\bibfnamefont {I.}~\bibnamefont
  {Skarmoutsos}}, \ and\ \bibinfo {author} {\bibfnamefont {M.}~\bibnamefont
  {Masia}},\ }\href@noop {} {\bibfield  {journal} {\bibinfo  {journal} {J.
  Chem. Theor. Comput.}\ }\textbf {\bibinfo {volume} {5}},\ \bibinfo {pages}
  {1449} (\bibinfo {year} {2009})}\BibitemShut {NoStop}%
\bibitem [{\citenamefont {{van Leeuwen}}\ and\ \citenamefont
  {Baerends}(1994)}]{baerends}%
  \BibitemOpen
  \bibfield  {author} {\bibinfo {author} {\bibfnamefont {R.}~\bibnamefont {{van
  Leeuwen}}}\ and\ \bibinfo {author} {\bibfnamefont {E.~J.}\ \bibnamefont
  {Baerends}},\ }\href@noop {} {\bibfield  {journal} {\bibinfo  {journal}
  {Phys. Rev. A}\ }\textbf {\bibinfo {volume} {49}},\ \bibinfo {pages} {2421}
  (\bibinfo {year} {1994})}\BibitemShut {NoStop}%
\bibitem [{\citenamefont {Niu}\ \emph {et~al.}(2011)\citenamefont {Niu},
  \citenamefont {Tan},\ and\ \citenamefont {Ichiye}}]{ichiye}%
  \BibitemOpen
  \bibfield  {author} {\bibinfo {author} {\bibfnamefont {S.}~\bibnamefont
  {Niu}}, \bibinfo {author} {\bibfnamefont {M.-L.}\ \bibnamefont {Tan}}, \ and\
  \bibinfo {author} {\bibfnamefont {T.}~\bibnamefont {Ichiye}},\ }\href@noop {}
  {\bibfield  {journal} {\bibinfo  {journal} {J. Chem. Phys.}\ }\textbf
  {\bibinfo {volume} {134}},\ \bibinfo {pages} {134501} (\bibinfo {year}
  {2011})}\BibitemShut {NoStop}%
\bibitem [{\citenamefont {Tsiper}\ and\ \citenamefont {Burke}(2004)}]{burke1}%
  \BibitemOpen
  \bibfield  {author} {\bibinfo {author} {\bibfnamefont {E.~V.}\ \bibnamefont
  {Tsiper}}\ and\ \bibinfo {author} {\bibfnamefont {K.}~\bibnamefont {Burke}},\
  }\href@noop {} {\bibfield  {journal} {\bibinfo  {journal} {J. Chem. Phys.}\
  }\textbf {\bibinfo {volume} {120}},\ \bibinfo {pages} {1153} (\bibinfo {year}
  {2004})}\BibitemShut {NoStop}%
\bibitem [{\citenamefont {Tsiper}(2005)}]{burke2}%
  \BibitemOpen
  \bibfield  {author} {\bibinfo {author} {\bibfnamefont {E.~V.}\ \bibnamefont
  {Tsiper}},\ }\href@noop {} {\bibfield  {journal} {\bibinfo  {journal} {Phys.
  Rev. Lett.}\ }\textbf {\bibinfo {volume} {94}},\ \bibinfo {pages} {013204}
  (\bibinfo {year} {2005})}\BibitemShut {NoStop}%
\bibitem [{\citenamefont {Paricaud}\ \emph {et~al.}(2005)\citenamefont
  {Paricaud}, \citenamefont {Predota}, \citenamefont {Chialvo},\ and\
  \citenamefont {Cummings}}]{cummings-w}%
  \BibitemOpen
  \bibfield  {author} {\bibinfo {author} {\bibfnamefont {P.}~\bibnamefont
  {Paricaud}}, \bibinfo {author} {\bibfnamefont {M.}~\bibnamefont {Predota}},
  \bibinfo {author} {\bibfnamefont {A.~A.}\ \bibnamefont {Chialvo}}, \ and\
  \bibinfo {author} {\bibfnamefont {P.~T.}\ \bibnamefont {Cummings}},\
  }\href@noop {} {\bibfield  {journal} {\bibinfo  {journal} {J. Chem. Phys.}\
  }\textbf {\bibinfo {volume} {122}},\ \bibinfo {pages} {244511} (\bibinfo
  {year} {2005})}\BibitemShut {NoStop}%
\bibitem [{\citenamefont {Halliwell}\ and\ \citenamefont
  {Nyburg}(1963)}]{halliwell}%
  \BibitemOpen
  \bibfield  {author} {\bibinfo {author} {\bibfnamefont {H.~F.}\ \bibnamefont
  {Halliwell}}\ and\ \bibinfo {author} {\bibfnamefont {S.~C.}\ \bibnamefont
  {Nyburg}},\ }\href@noop {} {\bibfield  {journal} {\bibinfo  {journal} {Trans.
  Faraday Soc.}\ }\textbf {\bibinfo {volume} {59}},\ \bibinfo {pages} {1126}
  (\bibinfo {year} {1963})}\BibitemShut {NoStop}%
\bibitem [{\citenamefont {Marcus}(1987)}]{marcus_tatb}%
  \BibitemOpen
  \bibfield  {author} {\bibinfo {author} {\bibfnamefont {Y.}~\bibnamefont
  {Marcus}},\ }\href@noop {} {\bibfield  {journal} {\bibinfo  {journal} {J.
  Chem. Soc. Faraday Trans.}\ }\textbf {\bibinfo {volume} {83}},\ \bibinfo
  {pages} {2985} (\bibinfo {year} {1987})}\BibitemShut {NoStop}%
\bibitem [{\citenamefont {Zhan}\ and\ \citenamefont
  {Dixon}(2001)}]{dixon_proton}%
  \BibitemOpen
  \bibfield  {author} {\bibinfo {author} {\bibfnamefont {C.-G.}\ \bibnamefont
  {Zhan}}\ and\ \bibinfo {author} {\bibfnamefont {D.~A.}\ \bibnamefont
  {Dixon}},\ }\href@noop {} {\bibfield  {journal} {\bibinfo  {journal} {J.
  Phys. Chem. A}\ }\textbf {\bibinfo {volume} {105}},\ \bibinfo {pages} {11534}
  (\bibinfo {year} {2001})}\BibitemShut {NoStop}%
\bibitem [{\citenamefont {Grossfield}\ \emph {et~al.}(2003)\citenamefont
  {Grossfield}, \citenamefont {Ren},\ and\ \citenamefont {Ponder}}]{agros03}%
  \BibitemOpen
  \bibfield  {author} {\bibinfo {author} {\bibfnamefont {A.}~\bibnamefont
  {Grossfield}}, \bibinfo {author} {\bibfnamefont {P.}~\bibnamefont {Ren}}, \
  and\ \bibinfo {author} {\bibfnamefont {J.~W.}\ \bibnamefont {Ponder}},\
  }\href@noop {} {\bibfield  {journal} {\bibinfo  {journal} {J. Am. Chem.
  Soc.}\ }\textbf {\bibinfo {volume} {125}},\ \bibinfo {pages} {15671}
  (\bibinfo {year} {2003})}\BibitemShut {NoStop}%
\bibitem [{\citenamefont {Lamoureux}\ and\ \citenamefont
  {Roux}(2006{\natexlab{b}})}]{glamo06}%
  \BibitemOpen
  \bibfield  {author} {\bibinfo {author} {\bibfnamefont {G.}~\bibnamefont
  {Lamoureux}}\ and\ \bibinfo {author} {\bibfnamefont {B.}~\bibnamefont
  {Roux}},\ }\href@noop {} {\bibfield  {journal} {\bibinfo  {journal} {J. Phys.
  Chem. B}\ }\textbf {\bibinfo {volume} {110}},\ \bibinfo {pages} {3308}
  (\bibinfo {year} {2006}{\natexlab{b}})}\BibitemShut {NoStop}%
\bibitem [{\citenamefont {Bryantsev}\ \emph {et~al.}(2008)\citenamefont
  {Bryantsev}, \citenamefont {Diallo},\ and\ \citenamefont {{Goddard
  III}}}]{goddard}%
  \BibitemOpen
  \bibfield  {author} {\bibinfo {author} {\bibfnamefont {Y.~S.}\ \bibnamefont
  {Bryantsev}}, \bibinfo {author} {\bibfnamefont {M.~S.}\ \bibnamefont
  {Diallo}}, \ and\ \bibinfo {author} {\bibfnamefont {W.~A.}\ \bibnamefont
  {{Goddard III}}},\ }\href@noop {} {\bibfield  {journal} {\bibinfo  {journal}
  {J. Phys. Chem. B}\ }\textbf {\bibinfo {volume} {112}},\ \bibinfo {pages}
  {9709} (\bibinfo {year} {2008})}\BibitemShut {NoStop}%
\bibitem [{\citenamefont {Buch}\ \emph {et~al.}(2007)\citenamefont {Buch},
  \citenamefont {Milet}, \citenamefont {Vacha}, \citenamefont {Jungwirth},\
  and\ \citenamefont {Devlin}}]{jungwirth-acid}%
  \BibitemOpen
  \bibfield  {author} {\bibinfo {author} {\bibfnamefont {V.}~\bibnamefont
  {Buch}}, \bibinfo {author} {\bibfnamefont {A.}~\bibnamefont {Milet}},
  \bibinfo {author} {\bibfnamefont {R.}~\bibnamefont {Vacha}}, \bibinfo
  {author} {\bibfnamefont {P.}~\bibnamefont {Jungwirth}}, \ and\ \bibinfo
  {author} {\bibfnamefont {J.~P.}\ \bibnamefont {Devlin}},\ }\href@noop {}
  {\bibfield  {journal} {\bibinfo  {journal} {Proc. Natl. Acad. Sci. USA}\
  }\textbf {\bibinfo {volume} {104}},\ \bibinfo {pages} {7342} (\bibinfo {year}
  {2007})}\BibitemShut {NoStop}%
\bibitem [{\citenamefont {Creux}\ \emph {et~al.}(2009)\citenamefont {Creux},
  \citenamefont {Lachaise}, \citenamefont {Graciaa}, \citenamefont {Beattie},\
  and\ \citenamefont {Djerdjev}}]{beattie}%
  \BibitemOpen
  \bibfield  {author} {\bibinfo {author} {\bibfnamefont {P.}~\bibnamefont
  {Creux}}, \bibinfo {author} {\bibfnamefont {J.}~\bibnamefont {Lachaise}},
  \bibinfo {author} {\bibfnamefont {A.}~\bibnamefont {Graciaa}}, \bibinfo
  {author} {\bibfnamefont {J.~K.}\ \bibnamefont {Beattie}}, \ and\ \bibinfo
  {author} {\bibfnamefont {A.~M.}\ \bibnamefont {Djerdjev}},\ }\href@noop {}
  {\bibfield  {journal} {\bibinfo  {journal} {J. Phys. Chem. B}\ }\textbf
  {\bibinfo {volume} {113}},\ \bibinfo {pages} {14146} (\bibinfo {year}
  {2009})}\BibitemShut {NoStop}%
\bibitem [{\citenamefont {Petersen}\ and\ \citenamefont
  {Saykally}(2008)}]{saykally-acid}%
  \BibitemOpen
  \bibfield  {author} {\bibinfo {author} {\bibfnamefont {P.~B.}\ \bibnamefont
  {Petersen}}\ and\ \bibinfo {author} {\bibfnamefont {R.~J.}\ \bibnamefont
  {Saykally}},\ }\href@noop {} {\bibfield  {journal} {\bibinfo  {journal}
  {Chem. Phys. Lett.}\ }\textbf {\bibinfo {volume} {458}},\ \bibinfo {pages}
  {244} (\bibinfo {year} {2008})}\BibitemShut {NoStop}%
\bibitem [{\citenamefont {Vaitheeswaran}\ and\ \citenamefont
  {Thirumalai}(2006)}]{vaitheeswaran_hydrophobic_2006}%
  \BibitemOpen
  \bibfield  {author} {\bibinfo {author} {\bibfnamefont {S.}~\bibnamefont
  {Vaitheeswaran}}\ and\ \bibinfo {author} {\bibfnamefont {D.}~\bibnamefont
  {Thirumalai}},\ }\href@noop {} {\bibfield  {journal} {\bibinfo  {journal} {J.
  Am. Chem. Soc.}\ }\textbf {\bibinfo {volume} {128}},\ \bibinfo {pages}
  {13490} (\bibinfo {year} {2006})}\BibitemShut {NoStop}%
\bibitem [{\citenamefont {Herce}\ \emph {et~al.}(2005)\citenamefont {Herce},
  \citenamefont {Perera}, \citenamefont {Darden},\ and\ \citenamefont
  {Sagui}}]{sagui05}%
  \BibitemOpen
  \bibfield  {author} {\bibinfo {author} {\bibfnamefont {D.~H.}\ \bibnamefont
  {Herce}}, \bibinfo {author} {\bibfnamefont {L.}~\bibnamefont {Perera}},
  \bibinfo {author} {\bibfnamefont {T.~A.}\ \bibnamefont {Darden}}, \ and\
  \bibinfo {author} {\bibfnamefont {C.}~\bibnamefont {Sagui}},\ }\href@noop {}
  {\bibfield  {journal} {\bibinfo  {journal} {J. Chem. Phys.}\ }\textbf
  {\bibinfo {volume} {122}},\ \bibinfo {pages} {024513} (\bibinfo {year}
  {2005})}\BibitemShut {NoStop}%
\bibitem [{\citenamefont {Jungwirth}\ and\ \citenamefont
  {Tobias}(2006)}]{jungwirth}%
  \BibitemOpen
  \bibfield  {author} {\bibinfo {author} {\bibfnamefont {P.}~\bibnamefont
  {Jungwirth}}\ and\ \bibinfo {author} {\bibfnamefont {D.~J.}\ \bibnamefont
  {Tobias}},\ }\href@noop {} {\bibfield  {journal} {\bibinfo  {journal} {Chem.
  Rev.}\ }\textbf {\bibinfo {volume} {106}},\ \bibinfo {pages} {1259} (\bibinfo
  {year} {2006})}\BibitemShut {NoStop}%
\bibitem [{\citenamefont {Kunz}\ \emph {et~al.}(2004)\citenamefont {Kunz},
  \citenamefont {{Lo Nostro}},\ and\ \citenamefont {Ninham}}]{wkunz04}%
  \BibitemOpen
  \bibfield  {author} {\bibinfo {author} {\bibfnamefont {W.}~\bibnamefont
  {Kunz}}, \bibinfo {author} {\bibfnamefont {P.}~\bibnamefont {{Lo Nostro}}}, \
  and\ \bibinfo {author} {\bibfnamefont {B.~W.}\ \bibnamefont {Ninham}},\
  }\href@noop {} {\bibfield  {journal} {\bibinfo  {journal} {Curr. Opin.
  Colloid Interface Sci.}\ }\textbf {\bibinfo {volume} {9}},\ \bibinfo {pages}
  {1} (\bibinfo {year} {2004})}\BibitemShut {NoStop}%
\bibitem [{\citenamefont {Ninham}\ and\ \citenamefont {{Lo
  Nostro}}(2010)}]{barrybook}%
  \BibitemOpen
  \bibfield  {author} {\bibinfo {author} {\bibfnamefont {B.~W.}\ \bibnamefont
  {Ninham}}\ and\ \bibinfo {author} {\bibfnamefont {P.}~\bibnamefont {{Lo
  Nostro}}},\ }\href@noop {} {\emph {\bibinfo {title} {Molecular Forces and
  Self Assembly}}}\ (\bibinfo  {publisher} {Cambridge, Cambridge},\ \bibinfo
  {year} {2010})\BibitemShut {NoStop}%
\bibitem [{\citenamefont {Chang}\ and\ \citenamefont
  {Dang}(2006)}]{chang_recent_2006}%
  \BibitemOpen
  \bibfield  {author} {\bibinfo {author} {\bibfnamefont {T.}~\bibnamefont
  {Chang}}\ and\ \bibinfo {author} {\bibfnamefont {L.}~\bibnamefont {Dang}},\
  }\href@noop {} {\bibfield  {journal} {\bibinfo  {journal} {Chem. Rev.}\
  }\textbf {\bibinfo {volume} {106}},\ \bibinfo {pages} {1305} (\bibinfo {year}
  {2006})}\BibitemShut {NoStop}%
\bibitem [{\citenamefont {Zhao}\ \emph {et~al.}(2010)\citenamefont {Zhao},
  \citenamefont {Rogers},\ and\ \citenamefont {Beck}}]{zhao10}%
  \BibitemOpen
  \bibfield  {author} {\bibinfo {author} {\bibfnamefont {Z.}~\bibnamefont
  {Zhao}}, \bibinfo {author} {\bibfnamefont {D.~M.}\ \bibnamefont {Rogers}}, \
  and\ \bibinfo {author} {\bibfnamefont {T.~L.}\ \bibnamefont {Beck}},\
  }\href@noop {} {\bibfield  {journal} {\bibinfo  {journal} {J. Chem. Phys.}\
  }\textbf {\bibinfo {volume} {132}},\ \bibinfo {pages} {014502} (\bibinfo
  {year} {2010})}\BibitemShut {NoStop}%
\bibitem [{\citenamefont {Collins}(2012)}]{collins-rev-12}%
  \BibitemOpen
  \bibfield  {author} {\bibinfo {author} {\bibfnamefont {K.~D.}\ \bibnamefont
  {Collins}},\ }\href@noop {} {\bibfield  {journal} {\bibinfo  {journal}
  {Biophys. Chem.}\ }\textbf {\bibinfo {volume} {167}},\ \bibinfo {pages} {43}
  (\bibinfo {year} {2012})}\BibitemShut {NoStop}%
\bibitem [{\citenamefont {{H\"{u}nenberger}}\ and\ \citenamefont
  {{McCammon}}(1999)}]{hunenberger-cont}%
  \BibitemOpen
  \bibfield  {author} {\bibinfo {author} {\bibfnamefont {P.~H.}\ \bibnamefont
  {{H\"{u}nenberger}}}\ and\ \bibinfo {author} {\bibfnamefont {J.~A.}\
  \bibnamefont {{McCammon}}},\ }\href@noop {} {\bibfield  {journal} {\bibinfo
  {journal} {J. Chem. Phys.}\ }\textbf {\bibinfo {volume} {110}},\ \bibinfo
  {pages} {1856} (\bibinfo {year} {1999})}\BibitemShut {NoStop}%
\bibitem [{\citenamefont {Reif}\ and\ \citenamefont
  {{H\"{u}nenberger}}(2011)}]{hunenberger_3}%
  \BibitemOpen
  \bibfield  {author} {\bibinfo {author} {\bibfnamefont {M.~M.}\ \bibnamefont
  {Reif}}\ and\ \bibinfo {author} {\bibfnamefont {P.~H.}\ \bibnamefont
  {{H\"{u}nenberger}}},\ }\href@noop {} {\bibfield  {journal} {\bibinfo
  {journal} {J. Chem. Phys.}\ }\textbf {\bibinfo {volume} {134}},\ \bibinfo
  {pages} {144103} (\bibinfo {year} {2011})}\BibitemShut {NoStop}%
\bibitem [{\citenamefont {Zhou}\ \emph {et~al.}(1988)\citenamefont {Zhou},
  \citenamefont {Stell},\ and\ \citenamefont {Friedman}}]{zsf}%
  \BibitemOpen
  \bibfield  {author} {\bibinfo {author} {\bibfnamefont {Y.}~\bibnamefont
  {Zhou}}, \bibinfo {author} {\bibfnamefont {G.}~\bibnamefont {Stell}}, \ and\
  \bibinfo {author} {\bibfnamefont {H.~L.}\ \bibnamefont {Friedman}},\
  }\href@noop {} {\bibfield  {journal} {\bibinfo  {journal} {J. Chem. Phys.}\
  }\textbf {\bibinfo {volume} {89}},\ \bibinfo {pages} {3836} (\bibinfo {year}
  {1988})}\BibitemShut {NoStop}%
\end{thebibliography}%

\end{document}